\documentclass[amsmath,amsfonts,superscriptaddress,prx,preprint,longbibliography]{revtex4-1}
\usepackage{graphicx}
\usepackage{epsfig}
\usepackage{bbm}
\usepackage{bm}
\usepackage{dcolumn}
\usepackage{amsmath}
\usepackage{amssymb}
\usepackage{color}

\newcommand{\beg}{\begin{equation}}
\newcommand{\en}{\end{equation}}

\newcommand{\br}{\mathbf r}

\newcommand \bel  {\begin{align}}
\newcommand \enl  {\end{align}}

\newcommand{\eps}{\epsilon}

\newcommand{\dg}{^\dagger}

\begin{document}

\title{Amplitude Higgs mode in superconductors with magnetic impurities}

\author{Yantao Li}
\address{Department of Physics, Kent State University, Kent, OH 44242, USA}

\author{Maxim Dzero}
\address{Department of Physics, Kent State University, Kent, OH 44242, USA}

\begin{abstract}
We study nonlinear response of conventional superconducting alloys with weak magnetic impurities to an external alternating electromagnetic field. In particular, we calculate a correction to superconducting order parameter $|\delta\Delta_{\Omega}|\exp(i\Omega t)$ up to the second order in external vector potential and show that frequency dependence of the order parameter amplitude has characteristic resonant shape with a maximum at frequency which is smaller than twice the magnitude of the pairing amplitude in equilibrium, $\Omega<2\Delta$, and at the same time exceeds the single-particle threshold energy. Our results suggest that in the presence of magnetic impurities the dynamics of the pairing amplitude in the collisionless regime will remain robust with respect to dissipative processes. We also evaluate the third harmonic contribution to the current as a function of the probe frequency and for various concentrations of magnetic impurities.
\end{abstract}


\date{\today}

\maketitle

\section{Introduction} Recent advances in state-of-the-art optical instruments and techniques 
have lead to an increased interest in the problems which 
focus on theoretical and experimental studies of various nonlinear responses in conventional and unconventional superconductors \cite{THz1,THz2,Shimano2012,Shimano2013,Shimano2014,THz3}. It is worth noting that while these developments must have been motivated, at least in part, by earlier theoretical discoveries such as stimulation of superconductivity by a microwave radiation (Eliashberg effect) and collisionless dynamics of the pairing amplitude in conventional Bardeen-Cooper-Schrieffer (BCS) superconductors \cite{Eliashberg1970,Klapwijk1977,Ivlev1973,VolkovKogan1973,Galaiko1972,Galperin1981}, the main conceptual motivation to make significant advances in this area of research has come from the realization that there exists a similarity between cosmology and condensed matter physics, specifically to superconductivity as well as other phases which exhibit well defined long-range order \cite{Demsar1999,Kaindl2000,Averitt2001,Leggett2005,Pashkin2010,Beck2011}. Indeed, the fully gapped amplitude mode in superconductors is similar to Higgs mode in quantum field theories \cite{Eckern1979,Varma1982,Eastham2011,Varma2014,Moore2017}. Therefore, by exploiting this similarity it becomes, in principle, feasible to probe the physics associated with the amplitude mode in a table-top experimental setup \cite{Shimano2020}.  

The excitation and propagation of the amplitude mode in superconductors are completely decoupled from the charge density fluctuations which are related to the phase fluctuations of the pairing field \cite{Moore2017}. On time scales which are short in comparison with the characteristic time scales for the single-particle relaxation processes, the dynamics of the pairing amplitude is described by the kinetic equations in which the collision integrals $I_{\textrm{e-e}}\propto\hbar/\tau_{\textrm{e-e}}$ and $I_{\textrm{e-ph}}\propto \hbar/\tau_{\textrm{e-ph}}$ which account for the electron-electron and electron-phonon scattering effects correspondingly, can be ignored \cite{VolkovKogan1973,Spivak2004,Burnett2005,Enolski2005,Enolski2005a,Altshuler2005,Levitov2006}. In other words, the dynamics of the amplitude mode is considered in the collisionless regime. Therefore, a problem of pairing amplitude dynamics becomes conceptually analogous to the one of the collisionless relaxation of an electric field in electronic plasma \cite{Landau1946,Kadomtsev1968}. Curiously, while the electric field in electronic plasma  attenuates exponentially fast after an initial perturbation (Landau damping), in conventional superconductors the amplitude mode asymptotes to a constant according to a power-law \cite{VolkovKogan1973,Gurarie2009,QReview2015}:
\beg\label{VolkovKogan}
|\Delta(t)|=\Delta_{\infty}\left(1+a\frac{\cos(2\Delta_\infty t+\pi/4)}{\sqrt{2\Delta_\infty t}}\right),
\en
where $a$ is some known parameter. The physical origin of this behavior has been understood using the exact solution to the problem of the BCS dynamics in a fermionic condensates \cite{Enolski2005,Enolski2005a,Altshuler2005,Yuzbashyan2006,Yuzbashyan2008}. Following the initial perturbation collective modes with frequencies $2\Omega_j=2(\eps_j^2+\Delta^2)^{1/2}$ are excited ($\eps_j$ are the roots of a certain nonlinear equation \cite{QReview2015}) and, in a complete analogy with the problem considered by Landau \cite{Landau1946,Kadomtsev1968,Eremin2023}, the dynamics of the pairing amplitude will be determined by a sum over excitation energies. In concert with the square-root anomaly in the density of states, this summation ultimately produces a power-law decay of the pairing amplitude, Eq. (\ref{VolkovKogan}), provided, of course, that the deviations from equilibrium are not too large \cite{QReview2015}.

It is important for our subsequent discussion to keep in mind that in the linear approximation, i.e. when the initial perturbation is weak (e.g. quenches of the pairing strength $g$ are of small magnitude, $|\delta g|\ll g$), $\Delta_\infty$ is equal to the value of the pairing amplitude in equilibrium, $\Delta$ \cite{SectionV}. Therefore, in the context of the pump-probe experiments one would expect that the resonant amplitude Higgs mode will be excited when the external frequency of the monochromatic field is tuned to $2\Omega_{\textrm{res}}=2\Delta$ \cite{Papenkort2007,Axt2009,Manske2014,Aoki2015,Kemper2015,Cea2016,Foster2017,Eremin2023}. Alternatively, when superconductor is in a state which carries a supercurrent, the amplitude Higgs mode will be excited at resonant frequency 
$\Omega_{\textrm{res}}=2\Delta$ \cite{Moore2017}.

And then there is a question whether the effects of potential disorder will affect the results we just discussed above for clean superconductors will be affected in any way. For zero-dimensional systems it is obvious that potential disorder will produce the renormalization of the single-particle energy levels and therefore will have no effect on the dynamics of the amplitude mode. In three dimensional systems the situation is more subtle. For a case of weak disorder Anderson theorem \cite{AndersonTheorem} guarantees that potential disorder should not have a significant effect on the dynamics and this conclusion should hold in both ballistic and diffusive regimes \cite{Silaev2019-Disorder,Seibold2021-Disorder,Haenel2021-Disorder,Yang2022-Disorder}. 
When disorder is strong enough to render the pairing interaction spatially inhomogeneous, Larkin and Ovchinnikov have shown \cite{LO-model} that in this case the inhomogeneities lead to the pair breaking and, at the mean-field level, their theory becomes analogous to the Abrikosov-Gor'kov theory of superconductors contaminated with paramagnetic impurities \cite{Dzero2021}. It is therefore expected that in this case amplitude dynamics may exhibit qualitatively different behavior from Eq. (1) and the results of the recent experiments on superconducting films near the superconductor-insulator transition \cite{Sherman2015-Disorder} seem to be in agreement with these observations, although the systematic theoretical analysis of these systems is inhibited by the fact that the ground states in strongly disordered superconducting films still remains very poorly understood \cite{KapitulnikRMP19,Dzero2023-SCQCP}.

Although the effects of potential disorder on an amplitude mode have already been studied, a question of what happens to the dynamics of the amplitude mode in superconducting alloys with magnetic impurities has not been addressed until very recently \cite{Dzero2023-Disorder}. This is quite surprising given how conceptually rich the problem of an interplay between conventional superconductivity and paramagnetic disorder really is (see e.g.  \cite{AG1961,Balatsky-RMP,Austin2000,Austin2001,Fominov2011,Feigel2013,Yerin-EPL2022,VityaG-2002} and references therein). The experimental progress in this direction is perhaps inhibited by the fact that it may be challenging to introduce the magnetic impurities in a control way such that their interplay with the dynamics of an amplitude mode can be probed already in the collisionless regime. 

One of the main results of \cite{Dzero2023-Disorder} consists in the following observation: when the relaxation time $\tau_s$ due to the scattering of conduction electrons on paramagnetic impurities is long enough so that the conditions $\tau_s\ll \tau_{\textrm{ee}}$ and $\zeta=1/\tau_s\Delta\ll1$ are met, after a quench of an arbitrarily small magnitude (linear approximation), the dynamics of the amplitude Higgs mode remains undamped 
\beg\label{MDzero}
|\Delta(t)|=\Delta\left\{1+\zeta\cos(\omega_s t+\pi/4)\right\}.
\en
The frequency of the Higgs mode oscillations is given by $\omega_s\approx2\Delta\sqrt{1-\zeta^2}<2\Delta$.
Clearly, Eq. (\ref{MDzero}) is very different from the Volkov-Kogan result, Eq. (\ref{VolkovKogan}), and it implies that scattering on paramagnetic impurities pushes the frequency of the Higgs mode below the minimum of the band of excitation energies $\Omega_j$ rendering it nondissipative. In passing we note, that in clean superconductors realization of the state with oscillating amplitude requires fairly large deviations from equilibrium \cite{Spivak2004,Levitov2006,Levitov2007,Dzero2008,Yuzbashyan2008}, which makes the result (\ref{MDzero}) - given that it appears already in the linear approximation - even more striking. We would like to emphasize that the results of Ref. \cite{Dzero2023-Disorder} are only valid in the perturbative regime $\zeta\ll 1$. Naturally, there are still questions which remain unanswered, such as the one about the fate of this non-dissipative amplitude mode when $\zeta\sim 1$, especially in the regime of gapless superconductivity \cite{AG1961}. Lastly, we note that similar findings have been recently in context of a problem when a clean superconductor is coupled to a strongly driven cavity \cite{Cavity-Higgs}, where external electromagnetic field in the cavity pushes the frequency of the Higgs mode below the gap edge and renders the order parameter dynamics to become periodic in time.

In Ref. \cite{Dzero2023-Disorder} the out-of-equilibrium dynamics in $s$-wave superconductor has been induced by a sudden, albeit small, change of the pairing strength. In this paper we consider a realistic situation when out-of-equilibrium dynamics is induced by an external electromagnetic $ac$-field and compute frequency dependence of the amplitude Higgs mode. We show that the resonance frequency at which this mode is excited is indeed smaller than $2\Delta$. At the same time, by evaluating the single particle density of states we demonstrate that it remains above the single particle threshold $\Delta_{\textrm{th}}=\Delta(1-\zeta^{2/3})^{3/2}$. These results are in general agreement with those of Ref. \cite{Dzero2023-Disorder}. In addition we compute the third harmonic contribution to the current in the pump-probe setup as a function of the probe frequency assuming the pump frequency has been tuned to a vicinity of the resonance amplitude mode frequency. We find that the largest contribution to the third harmonic is governed by the amplitude mode. We also find that the third harmonic contribution to the current is suppressed with an increase in magnetic scattering rate. We think that this particular result may shed some light on the physical origin of the energy scale corresponding to the resonant frequency of the amplitude mode.
We emphasize that our present findings are generally applicable for an arbitrary values of the dimensionless parameter $\zeta$. However, the effects associated with the formation of the Yu-Shiba-Rusinov bound states are not included in our forthcoming discussion and will be considered separately. 

\section{Basic equations}
In what follows we consider a disordered BCS superconductor in the diffusive limit $\Delta\ll 1/\tau$, where $\tau$ is the relaxation time due to scattering on potential impurities. It is clear that in the presence of the magnetic impurities this condition can be always fulfilled. At the same time we will assume that $\tau\ll \tau_s$. 

The central quantity for our analysis is the Green's function defined on the Keldysh contour:
\beg\label{GKeldysh}
\check{G}(t,t')=\left(\begin{matrix} \hat{G}^R(t,t') & \hat{G}^K(t,t') \\ 0 & \hat{G}^A(t,t')\end{matrix}\right).
\en
Each of component of the matrix function $\check{G}$ is a $4\times 4$ matrix and Nambu and spin subspaces \cite{Feigel2000,Austin2000}.
The Green's function (\ref{GKeldysh}) can be found by solving the Usadel equation for disordered superconductors, which corresponds to spatially homogeneous configuration of the $Q$-matrix at the saddle-point of the nonlinear $\sigma$-model \cite{Kamenev2009,Kamenev2011}:
\beg\label{Eq1Usadel}
i\left(\check{\Xi}_3\partial_t\check{G}+\partial_{t'}\check{G}\check{\Xi}_3\right)+\left[\check{\Delta},\check{G}\right]+\frac{i}{6\tau_s}\left[\left(\hat{\rho}_3\otimes\hat{\sigma}_i\right)\check{G}\left(\hat{\rho}_3\otimes\hat{\sigma}_i\right)\circ_,\check{G}\right]=-iD\left[{\check{\mathbf Q}}\check{G}{\check{\mathbf Q}}\circ_,\check{G}\right]. 
\en
Here $D=v_F^2\tau/3$ is the diffusion coefficient, ${\check{\mathbf Q}}(t)=(\hat{\gamma}_0\otimes\hat{\Xi}_3){\mathbf A}(t)$, ${\mathbf A}(t)$ is proportional to an external vector potential, $\check{\Xi}_3=\hat\gamma_0\otimes\hat{\Xi}_3$ is diagonal in Keldysh subspace,  $\hat{\gamma}_0$ is the unit Pauli matrix in the Keldysh space, $\hat{\Xi}_3=\hat{\rho}_3\otimes\hat{\sigma}_0$, $\left(\check{A}\circ\check{B}\right)(t,t')=\int dt_1\check{A}(t,t_1)\check{B}(t_1,t')$, $\hat{\rho}_n$ and $\hat{\sigma}_m$ ($n,m=1,2,3$) are the Pauli matrices acting in Nambu and spin subspaces correspondingly. 
Function $\check{G}$ must satisfy the normalization condition
\beg\label{Norm}
\check{G}\circ\check{G}=\check{{\mathbbm{1}}}
\en
and the third term in this equation should be understood as $\left[\check{\Delta},\check{G}\right]=\check{\Delta}(\br,t)\check{G}(\br;t,t')-\check{G}(\br;t,t')\check{\Delta}(\br,t')$, 
where matrix $\check{\Delta}(\br,t)=\Delta(\br,t)\left(\hat{\gamma}_0\otimes i\hat{\rho}_2\otimes\sigma_0\right)$ is diagonal in Keldysh space. 
The pairing field must be computed self-consistently from
\beg\label{Eq4Delta}
{\Delta}(t)=\frac{\pi\lambda}{2}\textrm{Tr}\left\{\left(\hat{\gamma}_1\otimes(\hat{\rho}_1-i\hat{\rho}_2)\otimes\hat{\sigma}_0\right)\check{G}(t,t)\right\}.
\en
Here $\lambda$ is the dimensionless pairing strength, $\hat{\gamma}_{1}$ is the first Pauli matrix acting in the Keldysh subspace. We would like to emphasize, that equation (\ref{Eq1Usadel}) has been found by performing an exact averaging over potential and magnetic disorder configurations. The only approximation that we have made is similar to one made in Refs.\cite{Austin2000,Austin2001} for the contribution from scattering on magnetic impurities and it is justified in the limit $\tau\ll \tau_s$. In other words, spatially homogeneous solution of (\ref{Eq1Usadel}) is applicable within the validity of the self-consistent Born approximation.
\subsection{Ground state}
The expressions for the components of $\check{G}(t,t')$ in the ground state are found by solving the Usadel equation (\ref{Eq1Usadel}) when external field is setzero:
\beg\label{Usadel0}
i\left(\check{\Xi}_3\partial_t\check{\cal G}+\partial_{t'}\check{\cal G}\check{\Xi}_3\right)+\left[\check{\Delta},\check{\cal G}\right]+\frac{i}{6\tau_s}\sum
\limits_{a=1}^3\left[\left(\hat{\rho}_3\otimes\hat{\sigma}_a\right)\check{\cal G}\left(\hat{\rho}_3\otimes\hat{\sigma}_a\right)\circ_,\check{\cal G}\right]=0.
\en
Performing the Fourier transform for the first term we find
\beg\label{First}
i\left(\check{\Xi}_3\partial_t\check{\cal G}+\partial_{t'}\check{\cal G}\check{\Xi}_3\right)=\int\frac{\eps d\eps}{2\pi}\left(\check{\Xi}_3\check{\cal G}_\eps-
\check{\cal G}_\eps\check{\Xi}_3\right)e^{-i\eps(t-t')}.
\en 
As it is well known, in equilibrium the Keldysh sub-block $\hat{\cal G}^K$ can always be chosen as:
\beg\label{GKParam}
\hat{\cal G}_\eps^K=\left(\hat{\cal G}_\eps^R-\hat{\cal G}_\eps^A\right)\tanh\left(\frac{\eps}{2T}\right).
\en
From the normalization condition (\ref{Norm}) we find
\beg\label{Norm2}
\hat{\cal G}_\eps^R\hat{\cal G}_\eps^R=\hat{\mathbbm 1}, \quad \hat{\cal G}_\eps^A\hat{\cal G}_\eps^A=\hat{\mathbbm 1}, \quad 
\hat{\cal G}_\eps^R\hat{\cal G}_\eps^K+\hat{\cal G}_\eps^K\hat{\cal G}_\eps^A=0.
\en
Below we will determine each of these Green's functions separately. 
\subsubsection{Retarded and advanced Green's functions}
Let us discuss an equation for the retarded matrix function $\hat{\cal G}_\eps^R$ first. From the form of the Usadel equation, we look for the solution for this function in the form:
\beg\label{AnsatzGR}
\hat{\cal G}_\eps^R=g_\eps^R\hat{\Xi}_3+f_\eps^R\hat{\Xi}_2.
\en
Here we introduced matrix $\hat{\Xi}_2=i\hat{\rho}_2\otimes\hat{\sigma}_0$.
Inserting this expression into equation (\ref{Usadel0}) yields
\beg\label{Usadel1}
\left(\eps+\frac{ig_\eps^R}{2\tau_{s}}\right)f_\eps^R
-\left(\Delta-\frac{if_\eps^R}{2\tau_{s}}\right)g_\eps^R=0.
\en
This equation is supplemented by the normalization condition
\beg\label{Normalize}
\left(g_\eps^R\right)^2-\left(f_\eps^R\right)^2=1.
\en
We can use the following standard parametrization for the functions $g_\eps^R=\cosh\theta_\eps$ and $f_\eps^R=\sinh\theta_\eps$.
Introducing
\beg\label{varaux}
\begin{split}
\tilde{\eps}=\eps+\frac{i}{2\tau_s}\cosh\theta_\eps, \quad \tilde{\Delta}_\eps=\Delta-\frac{i}{2\tau_s}\sinh\theta_\eps.
\end{split}
\en
we employ the normalization condition to write down the formal solution of (\ref{Usadel1}):
\beg\label{FormSol}
\cosh\theta_\eps=\frac{u_\eps}{\sqrt{u_\eps^2-1}}, \quad \sinh\theta_\eps=\frac{1}{\sqrt{u_\eps^2-1}}, \quad u_\eps=\frac{\tilde{\eps}}{\tilde{\Delta}_\eps}.
\en
Note that in the limit $\eps\gg\Delta$, it is implied that $u_\eps^{R(A)}=\pm\textrm{sign}\eps$. 

\begin{figure*}[t]
\includegraphics[width=0.475\linewidth]{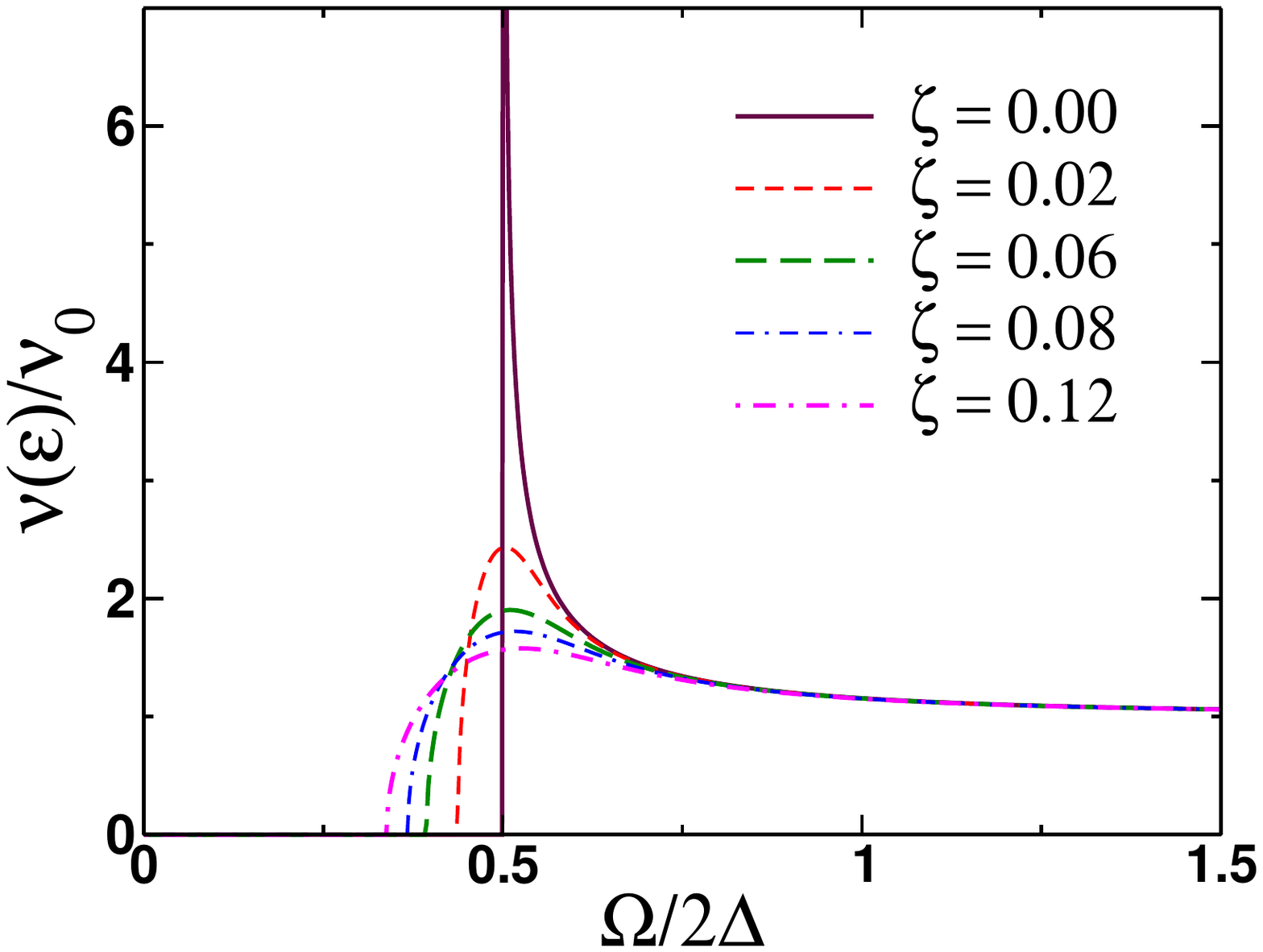}
\caption{Single particle density of states per spin as a function of energy evaluated for various values of the dimensionless parameter $\zeta=1/\tau_s\Delta$. }
\label{Fig-DOS}
\end{figure*}

Equations (\ref{FormSol}) is not a solution yet but just another parametrization of the Green's functions (\ref{AnsatzGR}). The actual solution of the Usadel equation determines the dependence of $g_\eps^R$ and $f_\eps^R$ on energy $\eps$, hence $u$ is a function of $\eps$ as well. The equation which allows one to compute the dependence of $u_\eps$ on $\eps$ reads:
\beg\label{Eq4ueps}
u_\eps\left(1-\frac{1}{\tau_s\Delta}\frac{1}{\sqrt{1-u_\eps^2}}\right)=\frac{\eps}{\Delta}.
\en
Thus, in what follows, we assume that the solution of equation (\ref{Eq4ueps}) is known and will work with the retarded and advanced Green's functions:
\beg\label{GRAFin}
\begin{split}
\hat{\cal G}_\eps^R=&\left(\frac{u_\eps}{\sqrt{u_\eps^2-1}}\hat{\Xi}_3+\frac{1}{\sqrt{u_\eps^2-1}}\hat{\Xi}_2\right)\textrm{sign}(\eps), \\
\hat{\cal G}_\eps^A=&-\hat{\Xi}_3\left(\hat{\cal G}_\eps^R\right)\dg\hat{\Xi}_3=-\frac{\overline{u}_\eps\textrm{sign}(\eps)}{\sqrt{\overline{u}_\eps^2-1}}\hat{\Xi}_3-\frac{\textrm{sign}(\eps)}{\sqrt{\overline{u}_\eps^2-1}}\hat{\Xi}_2.
\end{split}
\en
Here $\overline{u}_\eps=u_\eps^*$ implies a complex conjugation. Equation (\ref{Eq4ueps}) can be easily solved which allows one to compute the single particle density of states (DOS) per spin
\beg\label{DOS}
\frac{\nu(\eps)}{\nu_0}=\textrm{Re}\frac{u_\eps}{\sqrt{u_\eps^2-1}},
\en
where $\nu_0$ is the electron DOS per spin projection at the Fermi level in the normal state. We present the plots of $\nu(\eps)$ for various values of $\zeta=1/\tau_s\Delta$ in Fig. \ref{Fig-DOS}.
%%%%%%%%%%%%% Fig: DOS %%%%%%%%%%%%%%%%%
%%%%%%%%%%%%%%%%%%%%%%%%%%%%%%

\subsection{Application of an external electromagnetic field}
Having computed the Green's function in the ground state, we now look for the correction to the Green's function due to an application of external field.  We represent the external vector potential as a superposition of two monochromatic waves
\beg\label{ACfield}
{\mathbf A}(t)={\mathbf A}_{\Omega_1}e^{i\Omega_1t}+{\mathbf A}_{\Omega_2}e^{i\Omega_2t}+\textrm{c.c.}
\en
Our calculation will follow closely the avenue of Refs. \cite{Moore2017,Eremin2023}.
Specifically, we consider a correction to the Green's function $\check{\cal G}_\eps$: 
\beg\label{Gcorr}
\check{G}(\eps,\eps')=2\pi\check{\cal G}_\eps\delta(\eps-\eps')+\check{g}_1(\eps,\eps')
\en
and the correction to the order parameter $\check{\Delta}(t)=\check{\Delta}+\check{\Delta}_1(t)$.
The value of the unperturbed order parameter $\check{\Delta}$ must be computed self-consistently using equation (\ref{Eq4Delta}). Also, from the normalization condition it follows that the components of $\check{g}_1$ must satisfy
\beg\label{LinNorm}
\check{\cal G}_\eps \check{g}_1(\eps,\eps')+\check{g}_1(\eps,\eps')\check{\cal G}_{\eps'}=0.
\en

In the ground state we assumed that the order parameter is real. Under the action of the external field it may acquire an imaginary part. This is why the most general form of three Keldysh blocks in matrix Green's function $\check{g}_1$ must be of the form:
\beg\label{g1ansatz}
\hat{g}_1(\eps,\eps')=g_1(\eps,\eps')\hat{\Xi}_3+f_1(\eps,\eps')\hat{\Xi}_2.
\en
Note that due to the matrix form of (\ref{g1ansatz}), the corresponding matrix form of $\hat{\Delta}_1$ is the same as the one of $\hat{\Delta}$.

Now we go back to equation (\ref{Eq1Usadel}) and insert (\ref{Gcorr}) into the left hand side of that equation.
We keep the terms linear in $\check{g}_1$ and after performing the Fourier transformation we obtain
\beg\label{FourUsadel}
\begin{split}
&\left(\eps\check{\Xi}_3+\check{\Delta}\right)\check{g}_1(\eps,\eps')-\check{g}_1(\eps,\eps')\left(\eps'\check{\Xi}_3+\check{\Delta}\right)+\check{\Delta}_1(\eps'-\eps)\check{\cal G}_{\eps'}-\check{\cal G}_\eps\check{\Delta}_1(\eps'-\eps)\\&+\frac{i}{6\tau_s}\sum\limits_a\check{\Theta}_a
\check{\cal G}_\eps\check{\Theta}_a\check{g}_1(\eps,\eps')-\frac{i}{6\tau_s}\sum\limits_a\check{g}_1(\eps,\eps')\check{\Theta}_a\check{\cal G}_{\eps'}\check{\Theta}_a\\&=-2\pi iD\sum\limits_{\mu\nu}
\left(\check{\mathbf Q}_\nu\check{\cal G}_{\eps+\Omega_\nu}\check{\mathbf Q}_\mu\check{\cal G}_{\eps'}-\check{\cal G}_{\eps}\check{\mathbf Q}_\nu\check{\cal G}_{\eps'-\Omega_\mu}\check{\mathbf Q}_\mu\right)\delta\left(\eps'-\eps-\Omega_{\nu+\mu}\right),
\end{split}
\en
where $\check{\Theta}_a=\left(\hat{\gamma}_0\otimes\hat{\rho}_3\otimes\hat{\sigma}_a\right)$, $\Omega_{\nu,\mu}=\pm\Omega_{1,2}$ and $\Omega_{\nu+\mu}=\Omega_\nu+\Omega_\mu$.

Some re-arrangements of the few terms in this equation are in order. Let first look at the third and fourth terms in the left hand side of this equation: they have the same structure as the one in the right hand side. As it follows from the expression on the right hand side, we note that electromagnetic field will have an effect only when 
\beg\label{Resonance}
\eps'-\eps=\Omega_{\nu+\mu}.
\en
In other words, $\check{\Delta}_1(\eps'-\eps)$ is nonzero only when (\ref{Resonance}) holds. Then we use this observation to re-write the third and fourth terms as
\beg\label{ThirdFourth}
\check{\Delta}_1(\eps'-\eps)\check{\cal G}_{\eps'}-\check{\cal G}_\eps\check{\Delta}_1(\eps'-\eps)=2\pi\sum\limits_{\nu\mu}
\left[\check{\Delta}_1(\Omega_{\nu+\mu})\check{\cal G}_{\eps'}-\check{\cal G}_\eps\check{\Delta}_1(\Omega_{\nu+\mu})\right]\delta\left(\eps'-\eps-\Omega_{\nu+\mu}\right).
\en

The remaining terms in the left hand side can be simplified. Indeed, when $\tau_s\to\infty$, it is easy to see that $\left(\eps\check{\Xi}_3+\check{\Delta}\right)\propto\left[\check{\cal G}_\eps\right]_{\tau_s\to\infty}$. Taking these expressions into account, we can now re-write (\ref{FourUsadel}) as follows:
\beg\label{FourUsadel2}
\begin{split}
&\check{\Gamma}_\eps\check{g}_1(\eps,\eps')-\check{g}_1(\eps,\eps')\check{\Gamma}_{\eps'}=2\pi\sum\limits_{\nu\mu}\left[\check{\cal R}_{Q}(\eps,\eps')+\check{\cal R}_{\Delta}(\eps,\eps')\right]\delta\left(\eps'-\eps-\Omega_{\nu+\mu}\right).
\end{split}
\en
Here we introduced the following matrix functions:
\beg\label{DefRs}
\begin{split}
\check{\cal R}_{Q}(\eps,\eps')&=iD
\left(\check{\cal G}_{\eps}\check{\mathbf Q}_\nu\check{\cal G}_{\eps'-\Omega_\mu}\check{\mathbf Q}_\mu
-\check{\mathbf Q}_\nu\check{\cal G}_{\eps+\Omega_\nu}\check{\mathbf Q}_\mu\check{\cal G}_{\eps'}\right), \\
\check{\cal R}_{\Delta}(\eps,\eps')&=\check{\cal G}_\eps\check{\Delta}_1(\Omega_{\nu+\mu})-\check{\Delta}_1(\Omega_{\nu+\mu})\check{\cal G}_{\eps'},
\quad \check{\Gamma}_\eps=\eps\check{\Xi}_3+\check{\Delta}+\frac{i}{6\tau_s}\sum\limits_{a=1}^3\check{\Theta}_a
\check{\cal G}_\eps\check{\Theta}_a.
\end{split}
\en
As it is easy to check, in the limit $\tau_s\to\infty$ this equation coincides with the corresponding equations in \cite{Moore2017,Eremin2023}. 
Note that expression for $\check{\Gamma}_{\eps}$ has a non-zero Keldysh block which is not present in the case of potential disorder. 
We have to consider the solution of this equation for retarded, advanced and Keldysh sub-blocks separately. 
\subsubsection{Correction to the retarded and advanced Green's functions}
For the retarded and advanced blocks on the left hand side of (\ref{FourUsadel2}) it obtains
\beg\label{RAlhs}
\begin{split}
&\left[\check{\Gamma}_{\eps}\check{g}_1(\eps,\eps')-\check{g}_1(\eps,\eps')\check{\Gamma}_{\eps'}\right]^{R(A)}=
\left[(\zeta_\eps+\zeta_{\eps'})\hat{\cal G}_{\eps}\hat{g}_1(\eps,\eps')\right]^{R(A)}.
\end{split}
\en
Here we introduced functions 
\beg\label{DefEtas}
\zeta_\eps^{R}=\textrm{sign}(\eps)\tilde{\Delta}_\eps\sqrt{u_\eps^2-1}, \quad \zeta_\eps^{A}=-[\zeta_\eps^{(R)}]^*
\en
for they allow us to simplify the resulting expressions by employing the normalization condition.
It then follows
\beg\label{g1RA}
\hat{g}_1^{R(A)}(\eps,\eps')=2\pi\sum\limits_{\nu\mu}\left[\frac{
\hat{\cal G}_{\eps}\hat{\cal R}_{Q}(\eps,\eps')+\hat{\cal G}_{\eps}\hat{\cal R}_{\Delta}(\eps,\eps')}{\zeta_\eps+\zeta_{\eps'}}\right]^{R(A)}\delta\left(\eps'-\eps-\Omega_{\nu+\mu}\right).
\en
We would like to note that generally equation (\ref{Eq4ueps}) has two complex conjugate roots and so one needs to make sure that the root with the correct sign of the imaginary part is chosen such that  the retarded function in (\ref{DefEtas}) is analytic in upper half plane of the complex variable $\tilde{\eps}$.  
\subsubsection{Correction to the Keldysh Green's function}
Next we need to compute a correction to the remaining (Keldysh) block.
Correction to the Keldysh block of the Green's function is important, since it determines the correction to the order parameter (\ref{Eq4Delta}):
\beg\label{Delta1}
{\Delta}_{1}(t)=
\frac{\pi\lambda}{2}\int\frac{d\eps}{2\pi}\int\frac{d\eps'}{2\pi}\textrm{Tr}\left\{-\hat{\Xi}_2\hat{g}_1^K(\eps,\eps')\right\}e^{-i(\eps-\eps')t}.
\en
Function $\hat{g}_1^K(\eps,\eps')$ is itself proportional to ${\Delta}_{1}$, which will ultimately allow us to compute the pairing susceptibility. The frequency at which the susceptibility diverges determines the frequency of the amplitude (Higgs) mode $\omega_{\textrm{Higgs}}$. Therefore, we will be able to directly verify whether the $\omega_{\textrm{Higgs}}$ and $2\Delta$ are equal to each other or not. 

For the Keldysh component $\hat{g}_1^K(\eps,\eps')$ from (\ref{FourUsadel2}) we find:
\beg\label{g1K}
\begin{split}
\hat{N}_\eps^R\hat{g}_1^K(\eps,\eps')-\hat{g}_1^K(\eps,\eps')\hat{N}_{\eps'}^A&=2\pi
\sum\limits_{\nu\mu}\left[\hat{\cal R}_Q^K(\eps,\eps')+\hat{\cal R}_\Delta^K(\eps,\eps')\right]
\delta(\eps'-\eps-\Omega_{\nu+\mu})\\&+\hat{g}_1^R(\eps,\eps')\hat{\Lambda}_{\eps'}^K-
\hat{\Lambda}_\eps^K\hat{g}_1^A(\eps,\eps'),
\end{split}
\en
where $\check{N}_\eps=\eps\check{\Xi}_3+\check{\Delta}$ and $\check{\Lambda}_\eps=\check{\Gamma}_\eps-\check{N}_\eps$.
The last two terms in the right hand side of this equation appear explicitly due to scattering on paramagnetic impurities, since $\hat{\Lambda}_\eps^K\vert_{\tau_s\to \infty}=0$. 

In order to solve (\ref{g1K}) we again use the normalization condition (\ref{LinNorm}), which for the Keldysh components reads:
\beg\label{UseNormK}
\hat{\cal G}_{\eps}^R\hat{g}_1^K(\eps,\eps')+\hat{\cal G}_{\eps}^K\hat{g}_1^A(\eps,\eps')+\hat{g}_1^R(\eps,\eps')\hat{\cal G}_{\eps'}^K+\hat{g}_1^K(\eps,\eps')\hat{\cal G}_{\eps'}^A=0.
\en
We look for the solution of equation (\ref{g1K}) in the form:
\beg\label{g1KAnsatz}
\hat{g}_1^K(\eps,\eps')=\hat{g}_{1,\textrm{reg}}^K(\eps,\eps')+\hat{g}_{1,\textrm{an}}^K(\eps,\eps').
\en
The first dubbed as a regular term since it does not affect the single-particle distribution function in (\ref{g1KAnsatz})  is defined similarly to (\ref{GKParam}):
\beg\label{g1kreg}
\hat{g}_{1,\textrm{reg}}^K(\eps,\eps')=\hat{g}_{1}^R(\eps,\eps')n_{\eps'}-n_\eps\hat{g}_{1}^A(\eps,\eps'),
\en
where we are using the shorthand notation 
\beg\label{ne}
n_\eps=\tanh\left(\frac{\eps}{2T}\right).
\en
It is straightforward to verify that $\hat{g}_{1,\textrm{reg}}^K(\eps,\eps')$ satisfies the normalization condition (\ref{UseNormK}). On account of this fact, it is clear that $\hat{g}_{1,\textrm{an}}^K(\eps,\eps')$ must satisfy
\beg\label{NormKan}
\hat{\cal G}_{\eps}^R\hat{g}_{1,\textrm{an}}^K(\eps,\eps')+\hat{g}_{1,\textrm{an}}^K(\eps,\eps')\hat{\cal G}_{\eps'}^A=0.
\en
We now insert (\ref{g1KAnsatz}) into equation (\ref{g1K}) and after some algebra (see Appendix A) we find:
\beg\label{g1Kanom}
\hat{g}_{1,\textrm{an}}^K(\eps,\eps')=2\pi\sum\limits_{\nu\mu}\frac{\hat{\rho}_Q(\eps,\eps')+\hat{\rho}_\Delta(\eps,\eps')}{\zeta_\eps^R+\zeta_{\eps'}^A}\delta(\eps'-\eps-\Omega_{\nu+\mu}).
\en
The expressions for the matrix functions $\hat{\rho}_Q(\eps,\eps')$ and $\hat{\rho}_\Delta(\eps,\eps')$ are:
\beg\label{rhoDLTQ}
\begin{split}
\hat{\rho}_\Delta(\eps,\eps')&=\left[\hat{\cal G}_{\eps}^R\hat{\Delta}_1(\Omega_{\nu+\mu})\hat{\cal G}_{\eps'}^A-\hat{\Delta}_1(\Omega_{\nu+\mu})\right](n_{\eps'}-n_\eps),\\
\hat{\rho}_Q(\eps,\eps')&=iD\left[\hat{\cal G}_{\eps}^R\hat{\mathbf Q}_\nu\hat{\cal G}_{\eps+\Omega_\nu}^R\hat{\mathbf Q}_\mu\hat{\cal G}_{\eps'}^A-\hat{\mathbf Q}_\nu\hat{\cal G}_{\eps+\Omega_\nu}^R\hat{\mathbf Q}_\mu\right]
(n_{\eps'}-n_{\eps+\Omega_\nu})\\&-iD\left[\hat{\cal G}_\eps^R\hat{\mathbf Q}_\nu\hat{\cal G}_{\eps+\Omega_\nu}^A\hat{\mathbf Q}_\mu\hat{\cal G}_{\eps'}^A-\hat{\mathbf Q}_\nu\hat{\cal G}_{\eps+\Omega_\nu}^A\hat{\mathbf Q}_\mu\right](n_\eps-n_{\eps+\Omega_\nu}).
\end{split}
\en
Having computed $\hat{g}_1^K(\eps,\eps')$ we can directly insert it into the self-consistently equation and compute the resonant frequency of the amplitude Higgs mode.

\section{Amplitude Higgs mode}
We will analyze the self-consistency equation (\ref{Delta1}), which contains two contributions:
\beg\label{Delta1Main}
{\Delta}_{1}(t)=
\frac{\pi\lambda}{2}\int\limits_{-\infty}^\infty\frac{d\eps}{2\pi}\int\limits_{-\infty}^\infty\frac{d\eps'}{2\pi}\textrm{Tr}\left\{(-\hat{\Xi}_2)\left[\hat{g}_{1,\textrm{reg}}^K(\eps,\eps')
+\hat{g}_{1,\textrm{an}}^K(\eps,\eps')\right]\right\}e^{i(\eps'-\eps)t}.
\en
Without loss of generality, here we consider the case when 
\beg\label{EfetovChoice}
\Omega_1=0, \quad \Omega_2=\Omega. 
\en
This case is analogous to a setup in which a superconductor is prepared in a state which carries nonzero supercurrent. Thus, we will only need to focus on computing the Fourier component $\Delta_1(\Omega)=|\delta\Delta_\Omega|e^{i\Omega t}$. In the limit $\tau_s\to\infty$ 
amplitude $|\delta\Delta_\Omega|$ has a maximum at $\Omega_{\textrm{res}}=2\Delta$, which corresponds to the excitation of the amplitude Higgs mode \cite{Moore2017}. 

Using the expressions for the regular and anomalous contributions to the Keldysh component of the Green's functions from the previous section, the self-consistency equation for the Fourier component $\Delta_1(\Omega)$ can be cast into the following simple form
\beg\label{Self2Solve}
\Delta_1(\Omega)=2i\delta W_Q\left(\frac{B_{\textrm{reg}}(\Omega)+B_{\textrm{an}}(\Omega)}{A_{\textrm{reg}}(\Omega)+A_{\textrm{an}}(\Omega)}\right).
\en
Here $\delta W_Q=D{\mathbf A}_0{\mathbf A}_\Omega$. Functions $A_{\textrm{reg}}(\Omega)$ and $A_{\textrm{an}}(\Omega)$ are defined according to 
\beg\label{AB}
\begin{split}
A_{\textrm{reg}}(\Omega)&=\int\limits_{-\infty}^\infty{d\eps}\left(\frac{1+g_\eps^Rg_{\eps+\Omega}^R+f_\eps^Rf_{\eps+\Omega}^R}{\zeta_\eps^R+\zeta_{\eps+\Omega}^R}\right)n_{\eps+\Omega}\\&-\int\limits_{-\infty}^\infty{d\eps}\left(\frac{1+g_\eps^Ag_{\eps+\Omega}^A+f_\eps^Af_{\eps+\Omega}^A}{\zeta_\eps^A+\zeta_{\eps+\Omega}^A}-\frac{f_\eps^R-f_\eps^A}{\Delta}\right)n_\eps, \\
A_{\textrm{an}}(\Omega)&=\int\limits_{-\infty}^\infty\frac{\left(n_\eps-n_{\eps+\Omega}\right)}{\zeta_\eps^R+\zeta_{\eps+\Omega}^A}\left(1+g_\eps^Rg_{\eps+\Omega}^A+f_\eps^Rf_{\eps+\Omega}^A\right){d\eps}.
\end{split}
\en
The term $\propto (f_\eps^R-f_\eps^A)$ in the expression for $A_{\textrm{reg}}(\Omega)$ replaces the contribution from $1/\lambda$ by virtue of the self-consistency condition in equilibrium. The expression for $A_{\textrm{an}}(\Omega)$ involves a combination of retarded and advanced Green's functions and therefore have poles in both upper and lower half planes of complex variable $\eps$. This is not so for $A_{\textrm{reg}}(\Omega)$, in which all contributions containing retarded and advanced functions can be separated from each other. Therefore, in the expression for 
$A_{\textrm{reg}}(\Omega)$
we can reduce the integration over $\eps$ to the summation over the fermionic Matsubara frequencies $\omega_l=\pi T(2l+1)$ ($l=0,\pm1,...$) by using the series representation $\tanh x=\sum\limits_{l}2x/[\pi^2(l+1/2)^2+x^2]$. Subsequent integration in the upper or lower complex half plane with respect to complex $\eps$ then yields
\beg\label{Rule}
\int\limits_{-\infty}^\infty G^{R(A)}(\eps)\tanh\left(\frac{\eps}{2T}\right)d\eps =\pm 4\pi iT\sum\limits_{l=0}^\infty G(\pm i\omega_l).
\en
The corresponding expressions for the Green's functions are listed in the Appendix B.

Lastly, we also have found the following expressions for the functions $B_{\textrm{reg}}(\Omega)$ and $B_{\textrm{an}}(\Omega)$:
\beg\label{Breg}
\begin{split}
B_{\textrm{reg}}(\Omega)&=\int\limits_{-\infty}^\infty d\eps\left(\frac{g_\eps^R+g_{\eps+\Omega}^R}{\zeta_\eps^R+\zeta_{\eps+\Omega}^R}\right)(g_\eps^{R}f_{\eps+\Omega}^{R}+f_\eps^{R}g_{\eps+\Omega}^{R})n_{\eps+\Omega}\\&-\int\limits_{-\infty}^\infty d\eps\left(\frac{g_\eps^A+g_{\eps+\Omega}^A}{\zeta_\eps^A+\zeta_{\eps+\Omega}^A}\right)(g_\eps^{A}f_{\eps+\Omega}^{A}+f_\eps^{A}g_{\eps+\Omega}^{A})n_\eps, \\
B_{\textrm{an}}(\Omega)&=\int\limits_{-\infty}^\infty d\eps\frac{\left(n_\eps-n_{\eps+\Omega}\right)}{\zeta_\eps^R+\zeta_{\eps+\Omega}^A}\left(g_{\eps}^R+g_{\eps+\Omega}^A\right)\left(g_\eps^Rf_{\eps+\Omega}^A+f_\eps^Rg_{\eps+\Omega}^A\right).
\end{split}
\en
As we have discussed above, in order to compute the frequency dependence of $B_{\textrm{reg}}(\Omega)$ we convert the integral over $\eps$ into the summation over the fermionic Matsubara frequencies.  In passing we note that expressions (\ref{AB},\ref{Breg})  match the corresponding formulas in Refs. \cite{Moore2017,Eremin2023} and therefore we expect to recover their results in the limit $\zeta\to 0$. The results of the numerical calculation of the frequency dependence of the functions $A_{\textrm{reg}}(\Omega)+A_{\textrm{an}}(\Omega)$ and $B_{\textrm{reg}}(\Omega)+B_{\textrm{an}}(\Omega)$ can be found in Figs. \ref{FigADenom} and  \ref{FigBNumer} of Appendix B. 
%%%%%%%%%%%%% Fig: Abs[DLT1] %%%%%%%%%%%%%%%%%
\begin{figure*}[t]
\includegraphics[width=0.495\linewidth]{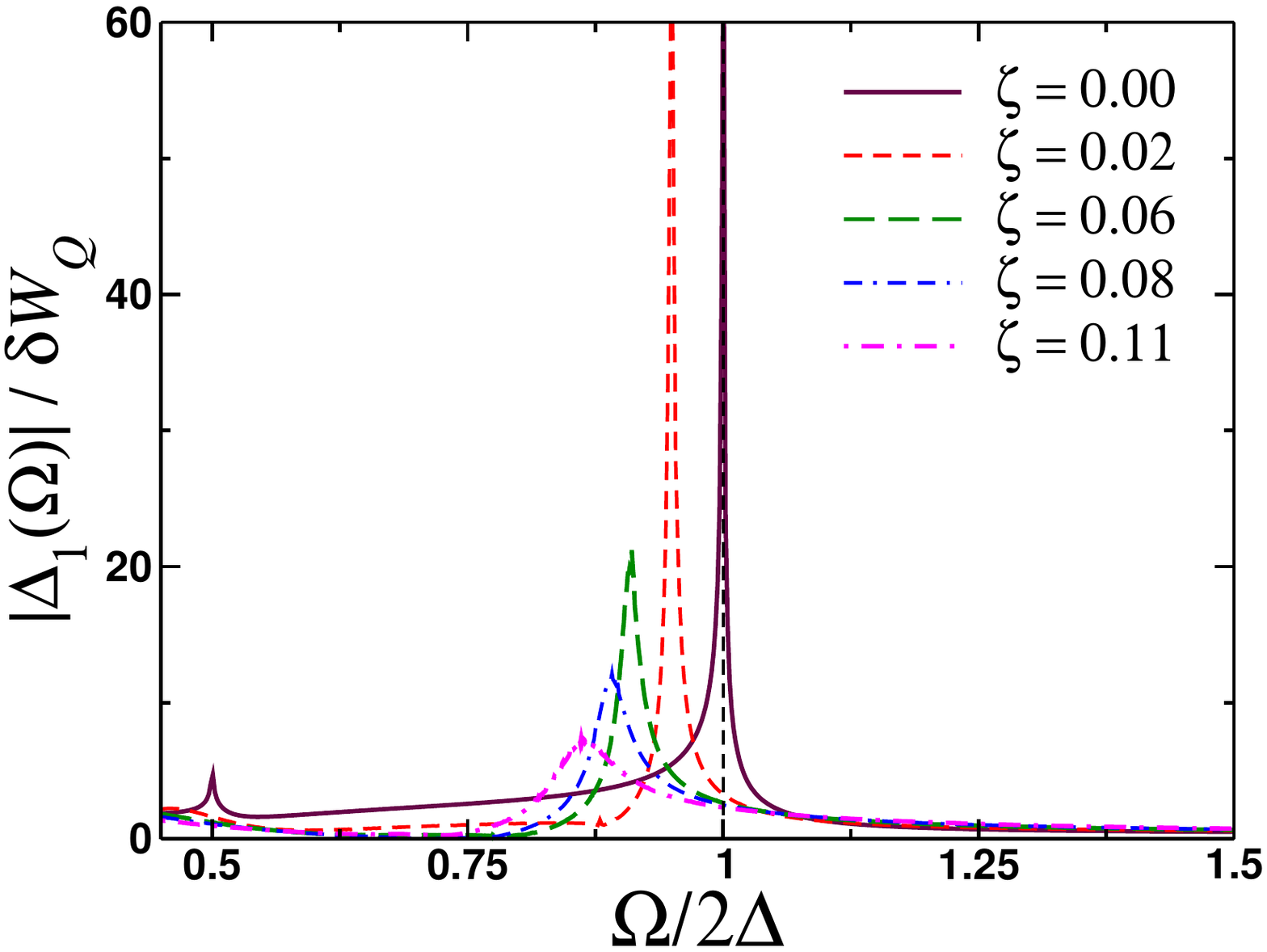}
\caption{Frequency dependence of the amplitude mode $\Delta_1(\Omega)$. The frequency $\Omega$ is shown in the units of the pairing amplitude $\Delta$ for various values of the dimensionless parameter $\zeta=1/\tau_s\Delta$. The amplitude mode has a maximum for $\zeta=0$ at $\Omega_{\textrm{res}}=2\Delta$. As we increase the value of the magnetic disorder parameter $\zeta$ the value of the resonant frequency shifts below $2\Delta$. Note that the amplitude of the resonant Higgs mode decreases with increasing strength of the magnetic disorder.}
\label{FigABSDLT1}
\end{figure*}
%%%%%%%%%%%%%%%%%%%%%%%%%%%%%%

Having computed these functions, we can now compute the amplitude of the resonant Higgs mode using equation (\ref{Self2Solve}). The results of the numerical calculation are presented in Fig. \ref{FigABSDLT1}. We immediately observe that with a small increase in the strength of magnetic scattering, the frequency of the resonant mode moves to the left, i.e. $\Omega_{\textrm{res}}(\zeta\not=0)<2\Delta$. This result qualitatively agrees with that of Ref. \cite{Dzero2023-Disorder}. We also see that the amplitude mode $|\Delta_1(\Omega)|$ decreases with an increase in $\zeta$. This latter result goes beyond the perturbative one of Ref. \cite{Dzero2023-Disorder}, where the amplitude of the periodic oscillations was proportional to $\zeta$. Therefore, one expects that the amplitude Higgs mode will be significantly suppressed before the gapless state is reached. The importance of our result $\Omega_{\textrm{res}}<2\Delta$ lies in fact that according to Ref. \cite{Dzero2023-Disorder} in this case the dynamics of the amplitude mode becomes dissipationless, i.e. $\Delta_1(t)$ will periodically vary in time on a time scale $t\ll \tau_{\textrm{e-e}}$. In order to show this explicitly within the confines of the present theoretical framework, we will have to determine the dynamics of the order parameter by solving the Usadel equation using (\ref{Gcorr}) as initial condition. This is an arduous task which we leave for the future studies. 

\section{Current induced by an external electromagnetic field}
In this Section we discuss the current induced by external electromagnetic radiation. Our main motivation is to get an insight into the origin of the shift in the resonance frequency from its value $2\Delta$ in a disordered superconductor without magnetic impurities. 
We consider an time-dependent external electric field
\beg\label{Efield}
{\mathbf E}(t)={\mathbf E}_\Omega\cos(\Omega t)+{\mathbf E}_{\omega_{\textrm{p}}}\cos(\omega_{\textrm{p}} t).
\en 
The first term here describes the 'pump field', ${\mathbf E}_\Omega$, while the second one is the 'probe field', ${\mathbf E}_{\omega_{\textrm{p}}}$. 
The vector potential ${\mathbf A}(t)$ has the same form as (\ref{Efield}) with the corresponding Fourier components given by:
\beg\label{Qmomentum}
{\mathbf A}_\Omega=-\frac{i{\mathbf E}_{\Omega}}{\Omega}, \quad {\mathbf A}_\omega=-\frac{i{\mathbf E}_{\omega}}{\omega}.
\en 
Here we use the units $\hbar=e=c=1$. The expression for the electric current can be compactly written as:
\beg\label{jnumain}
{\mathbf j}(\omega)=-{\cal Q}(\omega,\omega'){\mathbf A}_{\omega'},
\en
where ${\cal Q}(\omega,\omega')$ is the response kernel (we refer the reader to Appendix C for details):
\beg\label{QnnpMain}
\begin{split}
{\cal Q}(\omega,\omega')=\frac{\pi\sigma_D}{4i}\int\limits_{-\infty}^\infty \frac{d\eps}{2\pi}\int\limits_{-\infty}^\infty\frac{d\eps'}{2\pi}
&\textrm{Tr}\left\{\hat{\Xi}_3\hat{{G}}^R(\eps,\eps'-\omega')\hat{\Xi}_3\hat{{G}}^K(\eps',\eps+\omega)\right.\\&\left.+\hat{G}^K(\eps,\eps'-\omega')\hat{\Xi}_3\hat{{G}}^A(\eps',\eps+\omega)\hat{\Xi}_3\right\}.
\end{split}
\en
It is straightforward to verify that in the limit of very weak electromagnetic field we recover the familiar expression for the current \cite{Fominov2011}. 

\subsection{Third harmonic term in the current}
We will be mainly interested in the calculation of the third harmonic. One would generally expect that the third harmonic component of the kernel must display a feature (e. g. cusp) at $\omega_{\textrm{p}}=\Delta$ and we will interested whether this feature remain at $\Delta$ or will shift below $\Delta$ similar to the resonance frequency $\Omega_{\textrm{res}}$ discussed above.  
%%%%%%%%%%%%% Fig: Q3-reg 1,2 %%%%%%%%%%%%%%%%%
\begin{figure*}[t]
\includegraphics[width=0.475\linewidth]{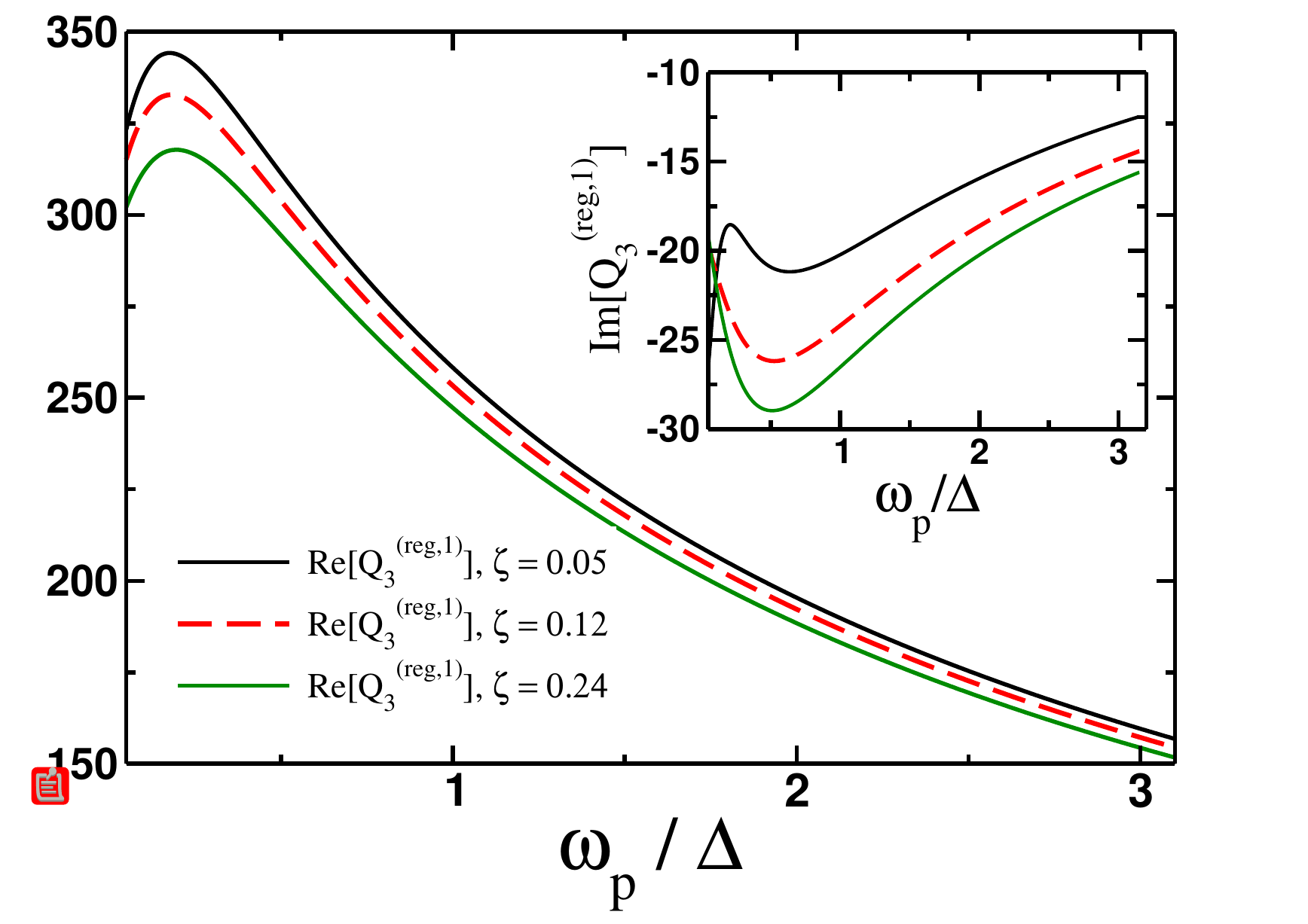}
\includegraphics[width=0.475\linewidth]{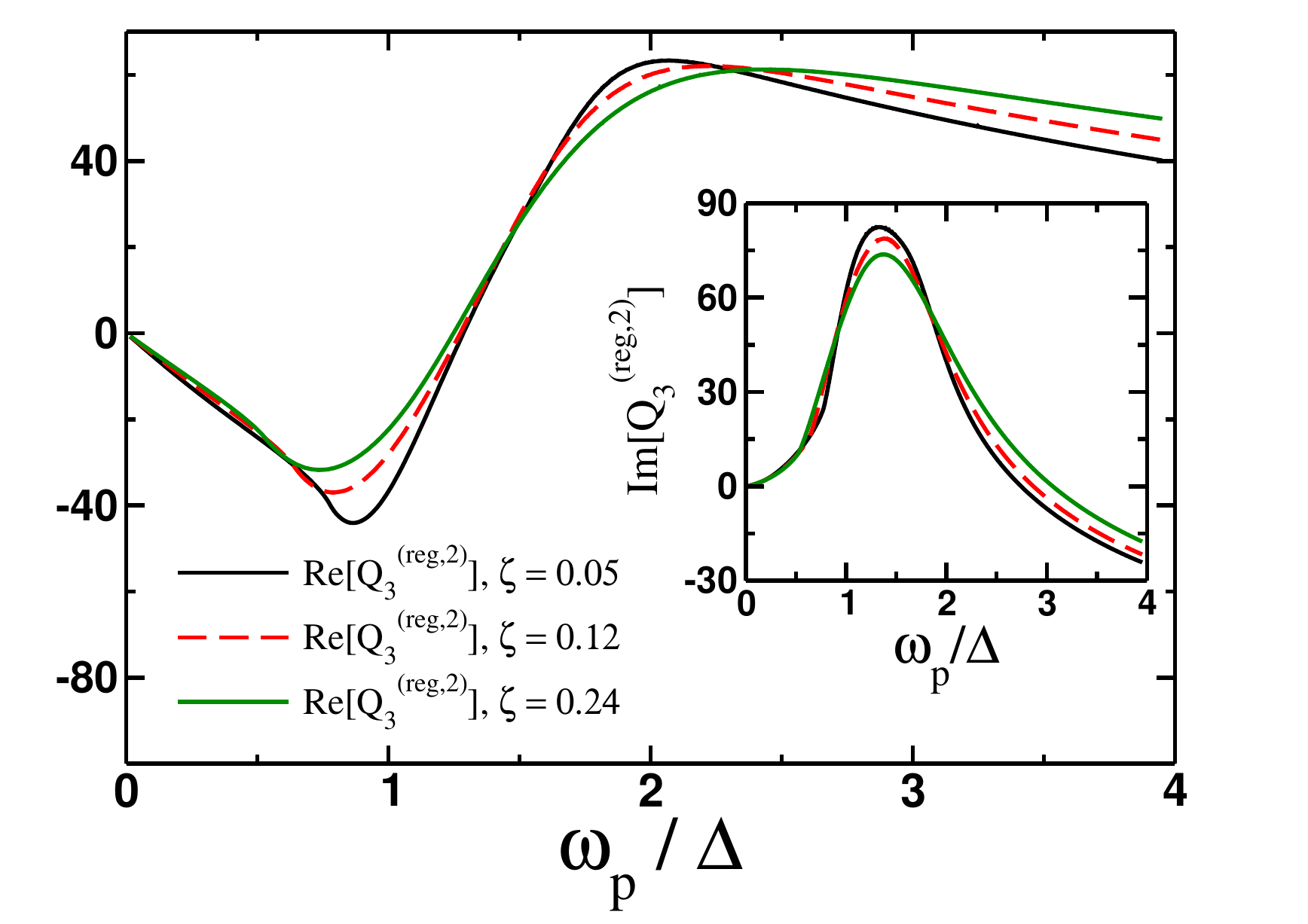}
\caption{Plots of the real (main) and imaginary (inset) parts of the functions ${\cal Q}_3^{(\textrm{reg1})}(\Omega,\omega_{\textrm{p}})$ (left panel) and ${\cal Q}_3^{(\textrm{reg2})}(\Omega,\omega_{\textrm{p}})$ (right panel) as a function of the probe frequency with the value of the pump frequency fixed to $\Omega=0.895\Delta$ for various values of the dimensionless parameter $\zeta$. Note that for the wide range of frequencies 
$|{\cal Q}_3^{(\textrm{reg1})}|\gg|{\cal Q}_3^{(\textrm{reg2})}|$. Both functions are given in the units of $\delta W_Q\sigma_D$. }
\label{Fig-Q3reg12}
\end{figure*}
%%%%%%%%%%%%%%%%%%%%%%%%%%%%%%

%%%%%%%%%%%%% Fig: Q3-an %%%%%%%%%%%%%%%%%
\begin{figure*}[t]
\includegraphics[width=0.475\linewidth]{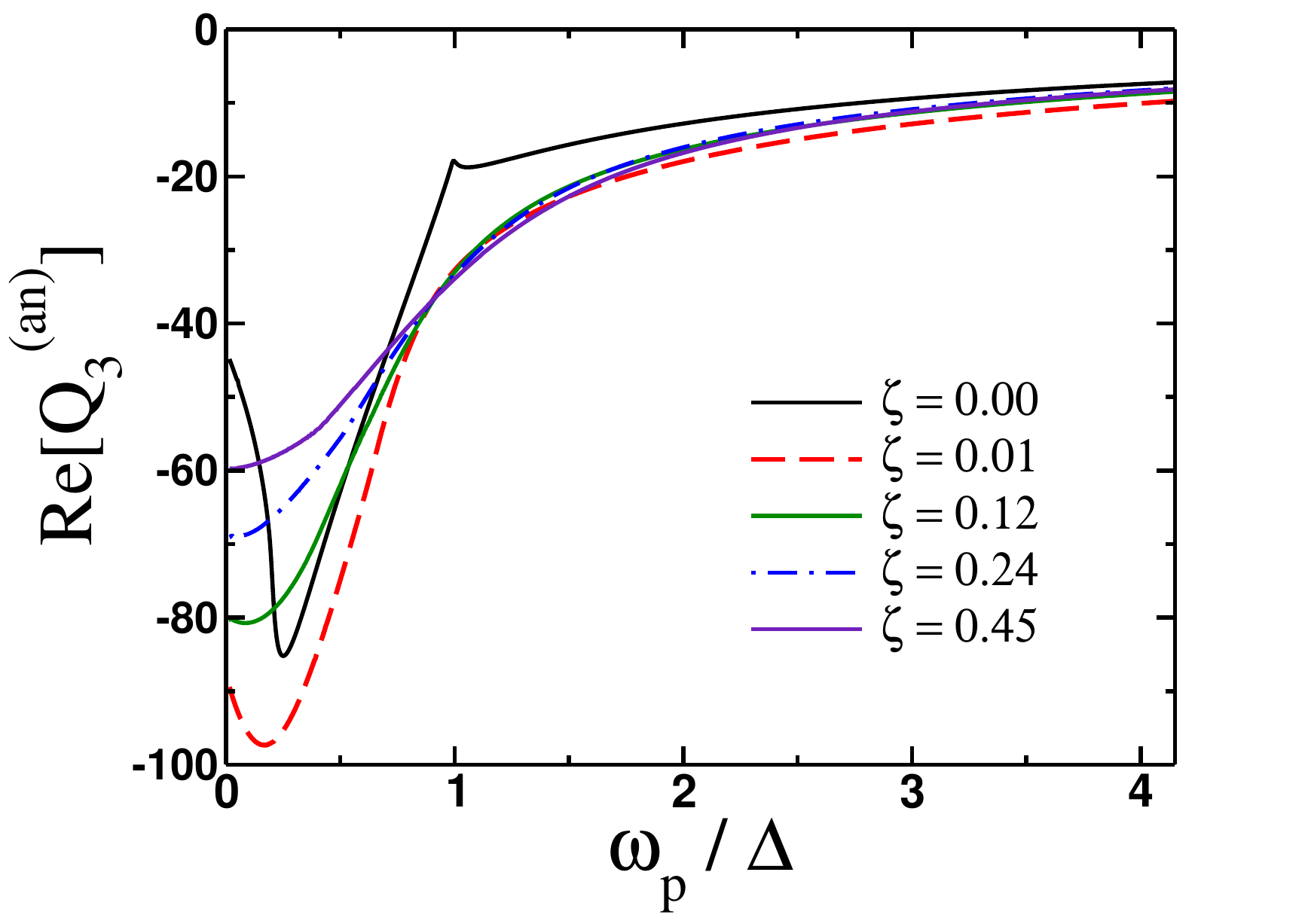}
\includegraphics[width=0.475\linewidth]{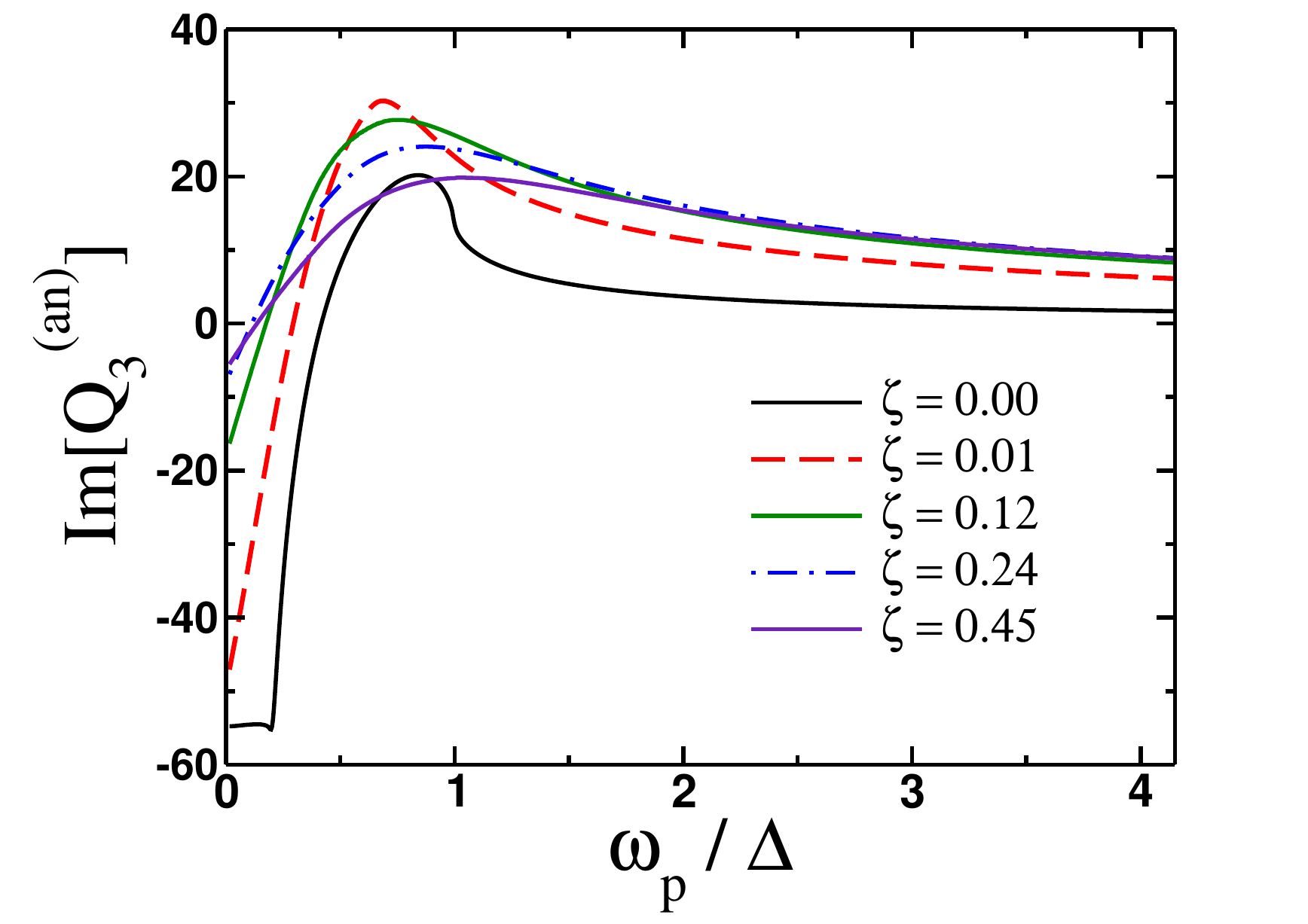}
\caption{Plots of the real (left panel) and imaginary (right panel) parts of the functions ${\cal Q}_{3}^{(\textrm{an})}(\Omega,\omega_{\textrm{p}})$ as functions of the probe frequency $\omega_{\textrm{p}}$ for several values of the dimensionless parameter $\zeta$. The value of the pump frequency has been fixed to $\Omega=0.895\Delta$.  Both functions are given in the units of $\delta W_Q\sigma_D$. For all the plots we have chosen $\Delta_1(2\Omega)=12\delta W_Q$.}
\label{Fig-Q3an}
\end{figure*}
%%%%%%%%%%%%%%%%%%%%%%%%%%%%%%

The response kernel for the third harmonic must be of the order of $O(A^2)$. It will be convenient to write it as a sum of three terms
\beg\label{ThreeKernel}
{\cal Q}_3={\cal Q}_3^{(\textrm{an})}+{\cal Q}_3^{(\textrm{reg1})}+{\cal Q}_3^{(\textrm{reg2})}.
\en
Here the first term is the anomalous one defined by $\hat{g}_{1,\textrm{an}}^K$ and it describes the effects associated with the non-equilibrium effects on the distribution function:
\beg\label{ThreeQsAn}
\begin{split}
{\cal Q}_3^{(\textrm{an})}(\omega,\omega')&=\frac{\sigma_D}{8i}\int\limits_{-\infty}^\infty
\textrm{Tr}\left\{\hat{g}_{1,\textrm{an}}^K(E,E+\omega-\omega')\left(\hat{\tilde{{\cal G}}}_{E-\omega'}^R+\hat{\tilde{{\cal G}}}_{E+\omega}^A\right)\right\}{dE}.
\end{split}
\en
The remaining two terms can be classified as the regular terms since they involve the distribution functions in equilibrium:
\beg\label{ThreeQsReg}
\begin{split}
{\cal Q}_3^{(\textrm{reg1})}(\omega,\omega')&=\frac{\sigma_D}{8i}\int\limits_{-\infty}^\infty
\textrm{Tr}\left\{\hat{g}_{1}^R(E,E+\omega-\omega')\left[\hat{\tilde{{\cal G}}}_{E-\omega'}^Rn_{E+\omega-\omega'}+\hat{\tilde{{\cal G}}}_{E+\omega}^Rn_{E+\omega}\right]\right\}{dE}\\&-\frac{\sigma_D}{8i}\int\limits_{-\infty}^\infty
\textrm{Tr}\left\{\hat{g}_{1}^A(E,E+\omega-\omega')\left[\hat{\tilde{{\cal G}}}_{E-\omega'}^An_{E-\omega'}+\hat{\tilde{{\cal G}}}_{E+\omega}^An_{E}\right]\right\}{dE}, \\
{\cal Q}_3^{(\textrm{reg2})}(\omega,\omega')&=\frac{\sigma_D}{8i}\int\limits_{-\infty}^\infty
\textrm{Tr}\left\{\hat{g}_{1}^R(E,E+\omega-\omega')\hat{\tilde{{\cal G}}}_{E+\omega}^A\right\}\left(n_{E+\omega-\omega'}-n_{E+\omega}\right){dE}\\&+\frac{\sigma_D}{8i}\int\limits_{-\infty}^\infty
\textrm{Tr}\left\{\hat{g}_{1}^A(E,E+\omega-\omega')\hat{\tilde{{\cal G}}}_{E-\omega'}^R\right\}\left(n_{E}-n_{E-\omega'}\right){dE}.
\end{split}
\en
We would like to remind the reader that $n_E$, Eq. (\ref{ne}), is not a single-particle Fermi distribution function, however it is related to it by 
$n_E=1-2n_F(E)$. The reason for considering two terms in (\ref{ThreeQsReg}) separately is purely technical: the integral  over energies in ${\cal Q}_3^{(\textrm{reg1})}$  can be converted into the summations over the fermionic Matsubara frequencies just like it has been done in the calculation of $\Delta_1(\Omega)$. Hence, we expect that this function should exhibit a monotonic behavior as a function of $\omega'$ for fixed $\omega$. 

We first proceed with the numerical calculation of the kernel ${\cal Q}_3$. In the expressions (\ref{rhoDLTQ}) we set $\Omega_\nu=\Omega_\mu=\Omega$ and $\Omega_{\nu+\mu}=2\Omega$. This implies that $\hat{g}_{1,\textrm{an}}^K$ is nonzero provided $\omega-\omega'=2\Omega$. Since by definition $\omega'=\omega_{\textrm{p}}$ (see Eq. (\ref{jnumain})) it follows that $\omega=2\Omega+\omega_{\textrm{p}}$.  We evaluate the dependence of 
${\cal Q}_3^{(\textrm{an})}$, ${\cal Q}_3^{(\textrm{reg1})}$ and ${\cal Q}_3^{(\textrm{reg2})}$ as functions of the probe frequency $\omega_{\textrm{p}}$ at low temperatures $T=10^{-5}\Delta$ and for $\Omega\sim \Delta$. The dependence of the complex functions ${\cal Q}_3^{(\textrm{reg1})}$ and ${\cal Q}_3^{(\textrm{reg2})}$ on the probe frequency in shown in Fig. \ref{Fig-Q3reg12} and the dependence of ${\cal Q}_3^{(\textrm{an})}$ is presented in Fig. \ref{Fig-Q3an}. Interestingly, we also observe that real part of ${\cal Q}_3^{(\textrm{reg1})}$ significantly exceeds the one of ${\cal Q}_3^{(\textrm{reg2})}$, while the imaginary parts are comparable to each other. This observation confirms our earlier expectations that the dominant contribution to both of these functions comes from the terms proportional to $\Delta_1$ \cite{Eremin2023}. In addition, we observe a cusp-like feature in the dependence of the Re$[{\cal Q}_3^{(\textrm{an})}]$ (and a 'weak discontinuity' in Im$[{\cal Q}_3^{(\textrm{an})}]$)  at $\omega_{\textrm{p}}\approx\Delta$ for $\zeta=0$, which shifts to smaller values and is almost completely smeared away at larger values of $\zeta$. A more detailed analysis of the third harmonic contribution to the current will be carried out when the experimental data will become available. 

\section{Discussion}
\begin{figure*}[t]
\includegraphics[width=0.495\linewidth]{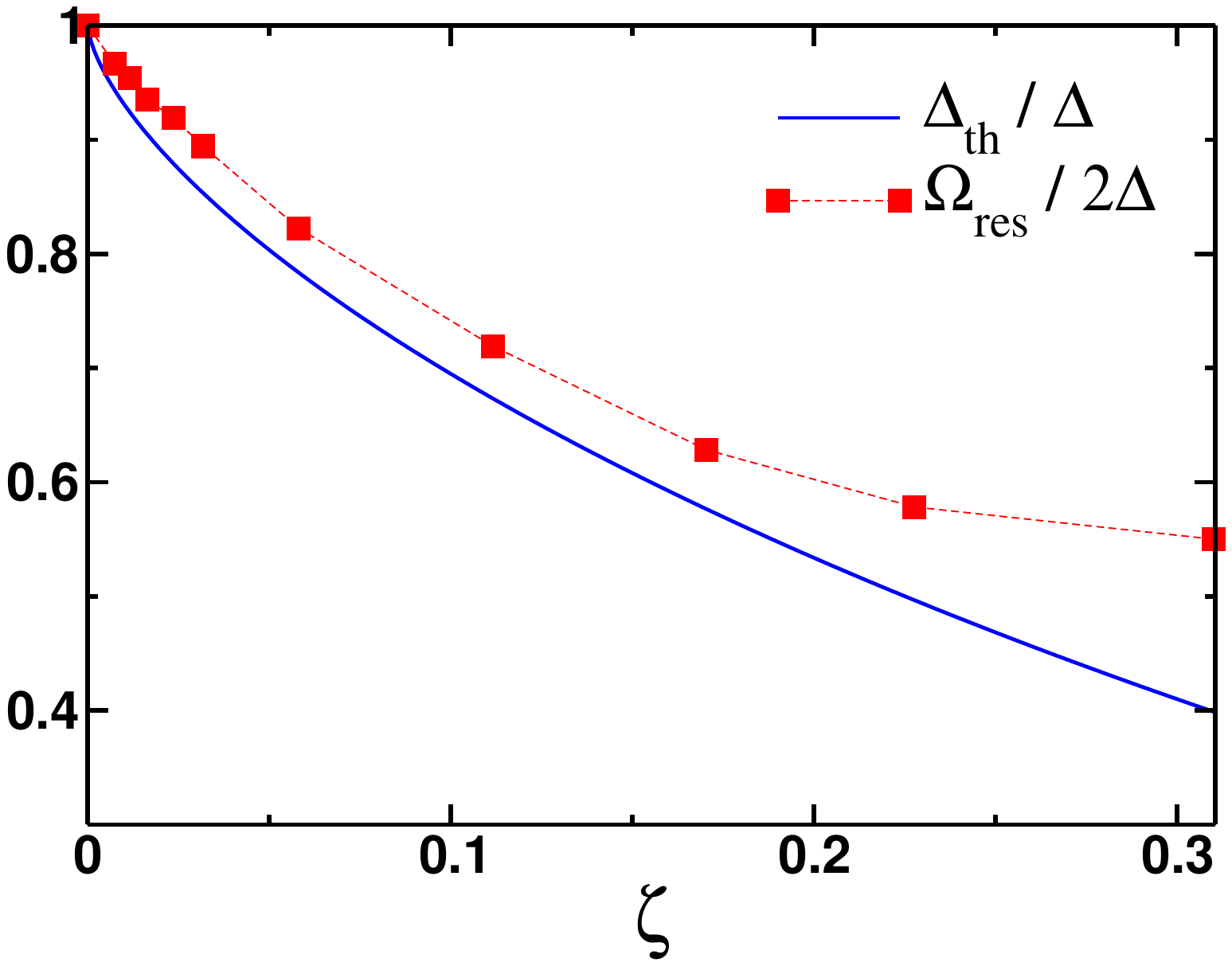}
\caption{The dependence of the single-particle threshold energy $\Delta_{\textrm{th}}=\Delta(1-\zeta^{2/3})^{3/2}$ and the resonant frequency of the amplitude Higgs mode $\Omega_{\textrm{res}}$ in the units of $2\Delta$. The frequency of the resonance mode remains above the single particle threshold energy, but below the pair excitation energy $2\Delta$. In the regime of the collisionless dynamics these results imply that the dynamics of the amplitude mode remains dissipationless on a time scale much shorter than the time scale for the two-particle collisions.}
\label{FigDLTth}
\end{figure*}

Our main finding - the reduction of the resonance frequency $\Omega_{\textrm{res}}$ below $2\Delta$ - requires further discussion.  
At first glance it seems that the decrease in the resonance frequency, Figs. \ref{FigABSDLT1} and \ref{FigDLTth}, along with the qualitatively similar finding of Ref. \cite{Dzero2023-Disorder} for the frequency of the Higgs mode $\omega_{\textrm{Higgs}}=2\Delta\sqrt{1-\zeta^2}$ emerge from the mathematics. We believe, however, that the approach we used in this paper allows one to give a clear physical interpretation of this result. Our calculation for the third harmonic of the electric current can be used to give more intuitive interpretation of the reduction $\Omega_{\textrm{res}}$. We first recall that in the diffusive superconductors even in the absence of magnetic disorder third harmonic generating current is mostly dominated by the amplitude Higgs mode \cite{Eremin2023}. Recall also that in the linear approximation the superfluid stiffness is directly proportional to the pairing amplitude and with an increase of magnetic scattering it decreases slower than the single-particle threshold energy, Fig. \ref{FigDLTth}. Therefore we are lead to conclude that the nonlinear suppression of the superfluid stiffness and the reduction in the frequency of the amplitude Higgs mode are two correlated effects, i.e. reduction of the superfluid stiffness through the nonlinear coupling to external electromagnetic field is reflected in the decrease of the resonant frequency. In this regard we suggest that in addition to two energy scales $\Delta_{\textrm{th}}(\zeta)$ and $\Delta(\zeta)$, for a complete description of a conventional superconductor with weak magnetic impurities one needs to consider an additional energy scale $\Omega_{\textrm{res}}(\zeta)$. Lastly, we mention that similar effect - reduction of the frequency of the amplitude Higgs mode below $2\Delta$ - has been discussed for various situations \cite{Cavity-Higgs,Bergeret2022} which are manifestly different from our mechanism and so the underlying physical processes responsible for this reduction are most likely different as well. 

However, we have to also mention that the dependence of $\Omega_{\textrm{res}}$ on $\zeta$ does not match the one of $\omega_{\textrm{Higgs}}$ on $\zeta$. The origin of this discrepancy is not clear to us at this point. In order to gain further insight one will have to compute the time dependence of the pairing amplitude $\Delta(t)$ after the electromagnetic pulse by solving directly the Usadel equation (\ref{Eq1Usadel}). From the form of the Usadel equation it is clear that the magnetic impurities do not lead to relaxation, therefore one will only needs to check if the magnetic impurities lead to the suppression of the dephasing processes which render the order parameter dynamics dissipationless. Given the fact that scattering on magnetic impurities leads to the smearing of the square-root singularity in the single particle density-of-states, Fig. \ref{Fig-DOS}, it is indeed likely that the dynamics of amplitude mode will exhibit periodic oscillations. 
Lastly, we have limited our discussion to the self-consistent Born approximation (SCBA) and did not consider the bound states which form on magnetic impurities. On a technical level, this requires a modification of the Usadel equation \cite{Marchetti2002,Fominov2011,Kharitonov2012}. We will consider the effects associated with the bound states - including the dynamics of the pairing amplitude and third harmonic of the current - in a separate publication. 

% Time dependence
% Bound states
% Possibility of experimental verification
%%%%%%%%%%%%% Fig: Abs[DLT1] %%%%%%%%%%%%%%%%%
%%%%%%%%%%%%%%%%%%%%%%%%%%%%%%
\section{Conclusions}
We have considered a problem of the nonlinear response of conventional (BCS) superconductors contaminated with weak magnetic impurities to external time-dependent electromagnetic field. Specifically, we have computed the resonant frequency of the amplitude mode and third harmonic contribution to electric current. Our main result is that the resonant frequency remains below the pair excitation threshold $2\Delta$ with an increase in scattering due to magnetic disorder. We attribute this shift to the nonlinear suppression of the superfluid density. Taken together with the results of Ref. \cite{Dzero2023-Disorder}, our present findings unambiguously suggest that the dynamics of the pairing amplitude should remain periodic in time. We have also found that with an increase in magnetic scattering  the third harmonic is suppressed along with the amplitude of the resonant mode. We attribute effect to the nonlinear suppression of the superfluid density. 

\section{Acknowledgements} We would like to thank Maxim Khodas, Alex Levchenko and Dima Pesin for their interest in this work and useful discussions. We gratefully acknowledge the financial support by the National Science Foundation grant NSF-DMR-2002795 (Y. L. and M. D.) This project was started during the Aspen Center of Physics 2023 Summer Program on "\emph{New Directions on Strange Metals in Correlated Systems}", which was supported by the National Science Foundation Grant No. PHY-2210452.

\begin{appendix}
\section{Derivation of the expression for $\hat{g}_{1,\textrm{an}}^K(\eps,\eps')$}
In this Appendix we provide the details on the derivation of the equation (\ref{g1Kanom}) in the main text. 
Our starting point is equation (\ref{g1K}) in the main text. 
Let us first consider terms which contain $\hat{g}_{1,\textrm{reg}}^K(\eps,\eps')$ only:
\beg\label{gregsimp1}
\begin{split}
&\hat{N}_\eps^R\hat{g}_{1,\textrm{reg}}^K(\eps,\eps')-\hat{g}_{1,\textrm{reg}}^K(\eps,\eps')\hat{N}_{\eps'}^A=\left(\hat{N}_\eps^R\hat{g}_1^R-\hat{g}_1^R\hat{N}_{\eps'}^A\right)n_{\eps'}-n_\eps\left(\hat{N}_\eps^R\hat{g}_1^A-\hat{g}_1^A\hat{N}_{\eps'}^A\right)\\&=\left(\hat{N}_\eps^R\hat{g}_1^R-\hat{g}_1^R\hat{N}_{\eps'}^R\right)n_{\eps'}-
n_\eps\left(\hat{N}_\eps^A\hat{g}_1^A-\hat{g}_1^A\hat{N}_{\eps'}^A\right)+g_1^R\left(\hat{N}_{\eps'}^R-\hat{N}_{\eps'}^A\right)n_{\eps'}\\&-n_\eps\left(\hat{N}_{\eps}^R-\hat{N}_{\eps}^A\right)g_1^A.
\end{split}
\en
If we now look at the equation (\ref{FourUsadel2}) and note that $\check{\Lambda}_\eps$ has only non-zero Keldysh component. We have
\beg\label{Replace1}
\begin{split}
&\left(\hat{N}_\eps^R\hat{g}_1^R-\hat{g}_1^R\hat{N}_{\eps'}^R\right)n_{\eps'}-
n_\eps\left(\hat{N}_\eps^A\hat{g}_1^A-\hat{g}_1^A\hat{N}_{\eps'}^A\right)\\&=2\pi\sum\limits_{\nu\mu}\left[\hat{\cal R}_Q^R(\eps,\eps')n_{\eps'}-n_\eps\hat{\cal R}_Q^A(\eps,\eps')+\hat{\cal R}_\Delta^R(\eps,\eps')n_{\eps'}-n_\eps\hat{\cal R}_\Delta^A(\eps,\eps')\right]\delta\left(\eps'-\eps-\Omega_{\nu+\mu}\right)\\&\equiv\hat{\cal P}^R(\eps,\eps')n_{\eps'}-n_\eps\hat{\cal P}^A(\eps,\eps').
\end{split}
\en
Here $\hat{\cal P}^{R(A)}(\eps,\eps')$ are used here to keep the expressions as compact as possible. Let us now consider the remaining two terms in (\ref{gregsimp1}):
\beg\label{Replace2}
\begin{split}
&\hat{g}_1^R(\eps,\eps')\left(\hat{N}_{\eps'}^R-\hat{N}_{\eps'}^A\right)n_{\eps'}-n_\eps\left(\hat{N}_{\eps}^R-\hat{N}_{\eps}^A\right)\hat{g}_1^A(\eps,\eps')\\&=\hat{g}_1^R(\eps,\eps')\left[\left({g}_{\eps'}^R-g_{\eps'}^A\right)\hat{\Xi}_3-
(f_{\eps'}^R-f_{\eps'}^A)\hat{\Xi}_2\right]\frac{in_{\eps'}}{2\tau_s}\\&-
\frac{in_{\eps}}{2\tau_s}\left[\left({g}_{\eps}^R-g_{\eps}^A\right)\hat{\Xi}_3-
(f_{\eps}^R-f_{\eps}^A)\hat{\Xi}_2\right]\hat{g}_1^A(\eps,\eps')=\hat{g}_1^R(\eps,\eps')\hat{\Lambda}_{\eps'}^K-\hat{\Lambda}_\eps^K\hat{g}_1^A(\eps,\eps')
\end{split}
\en
and we took into account formula (\ref{GKParam}) in the main text. Therefore, we have found that
\beg\label{g1regfin}
\begin{split}
\hat{N}_\eps^R\hat{g}_{1,\textrm{reg}}^K(\eps,\eps')-\hat{g}_{1,\textrm{reg}}^K(\eps,\eps')\hat{N}_{\eps'}^A&=
\hat{\cal P}^R(\eps,\eps')n_{\eps'}-n_\eps\hat{\cal P}^A(\eps,\eps')\\&+\hat{g}_1^R(\eps,\eps')\hat{\Lambda}_{\eps'}^K-\hat{\Lambda}_\eps^K\hat{g}_1^A(\eps,\eps').
\end{split}
\en
We now insert this expression into equation (\ref{g1K}). The terms which contain $\check{\Lambda}_\eps$ cancel out and it obtains:
\beg\label{g1Ka}
\begin{split}
\hat{N}_\eps^R\hat{g}_{1,\textrm{an}}^K(\eps,\eps')-\hat{g}_{1,\textrm{an}}^K(\eps,\eps')\hat{N}_{\eps'}^A&=
\hat{\cal P}^K(\eps,\eps')-\hat{\cal P}^R(\eps,\eps')n_{\eps'}+n_\eps\hat{\cal P}^A(\eps,\eps').
\end{split}
\en
Interestingly, this equation has the same form as the one for the case of nonmagnetic disorder. Let us now simplify the expression in the right hand side of the equation (\ref{g1Ka}):
\beg\label{PDLT}
\begin{split}
&\hat{\cal P}_\Delta^K(\eps,\eps')-\hat{\cal P}_\Delta^R(\eps,\eps')n_{\eps'}+n_\eps\hat{\cal P}_\Delta^A(\eps,\eps')\\&=
2\pi
\sum\limits_{\nu\mu}(n_\eps-n_{\eps'})
\left[\hat{\cal G}_\eps^R\hat{\Delta}_1(\Omega_{\nu+\mu})-\hat{\Delta}_1(\Omega_{\nu+\mu})\hat{\cal G}_{\eps'}^A\right]\delta\left(\eps'-\eps-\Omega_{\nu+\mu}\right).
\end{split}
\en
In passing we note that this result matches the corresponding expressions in Refs. \cite{Moore2017,Eremin2023}. For the remaining contribution we find
\beg\label{PQ}
\begin{split}
&\hat{\cal P}_{Q}^K(\eps,\eps')-\hat{\cal P}_{Q}^R(\eps,\eps')n_{\eps'}+n_\eps\hat{\cal P}_{Q}^A(\eps,\eps')\\&=
2\pi iD\sum\limits_{\nu\mu}\left[\hat{\mathbf Q}_\nu\hat{\cal G}_{\eps+\Omega_\nu}^R\hat{\mathbf Q}_\mu\hat{\cal G}_{\eps'}^A-\hat{\cal G}_\eps^R\hat{\mathbf Q}_\nu\hat{\cal G}_{\eps+\Omega_\nu}^R\hat{\mathbf Q}_\mu\right]
(n_{\eps'}-n_{\eps+\Omega_\nu})\delta(\eps'-\eps-\Omega_{\nu+\mu})\\&-2\pi iD\sum\limits_{\nu\mu}\left[\hat{\mathbf Q}_\nu\hat{\cal G}_{\eps+\Omega_\nu}^A\hat{\mathbf Q}_\mu\hat{\cal G}_{\eps'}^A-\hat{\cal G}_\eps^R\hat{\mathbf Q}_\nu\hat{\cal G}_{\eps+\Omega_\nu}^A\hat{\mathbf Q}_\mu\right](n_\eps-n_{\eps+\Omega_\nu})\delta(\eps'-\eps-\Omega_{\nu+\mu}),
\end{split}
\en
which is also in agreement with the results of Refs. \cite{Moore2017,Eremin2023}.
Finally, we represent
\beg\label{Rep4g1K}
\hat{N}_\eps^R\hat{g}_{1,\textrm{an}}^K(\eps,\eps')-\hat{g}_{1,\textrm{an}}^K(\eps,\eps')\hat{N}_{\eps'}^A=
(\zeta_\eps^R+\zeta_{\eps'}^A)\hat{\cal G}_\eps^R\hat{g}_1^K(\eps,\eps')
\en
and solve the resulting equation for $\hat{g}_{1,\textrm{an}}^K(\eps,\eps')$ by employing the normalization condition. This yields (\ref{g1Kanom}) in the main text.
%%%%%%%%%%%%% Fig: Areg+Aan %%%%%%%%%%%%%%%%%
\begin{figure*}[t]
\includegraphics[width=0.475\linewidth]{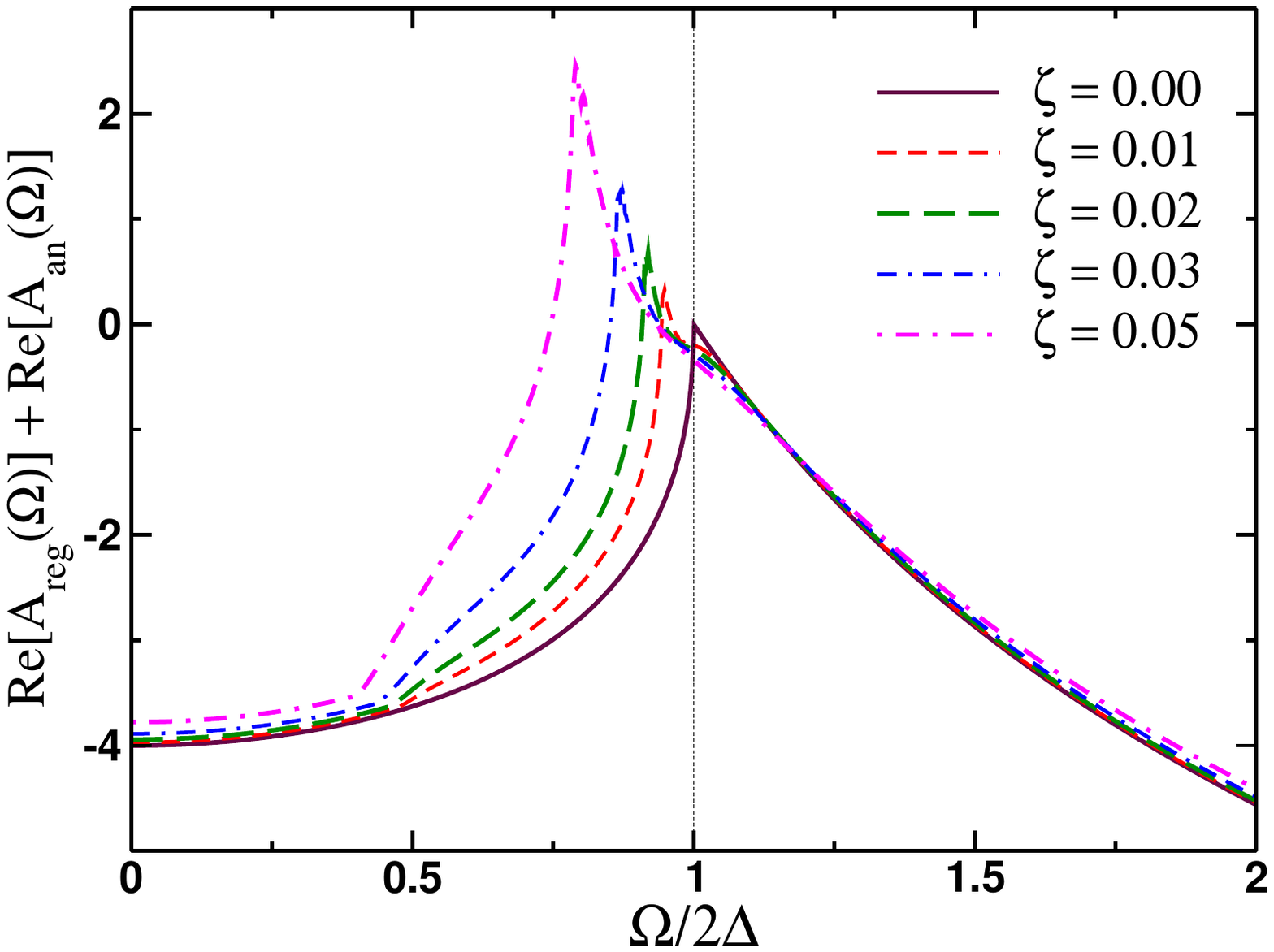}
\includegraphics[width=0.475\linewidth]{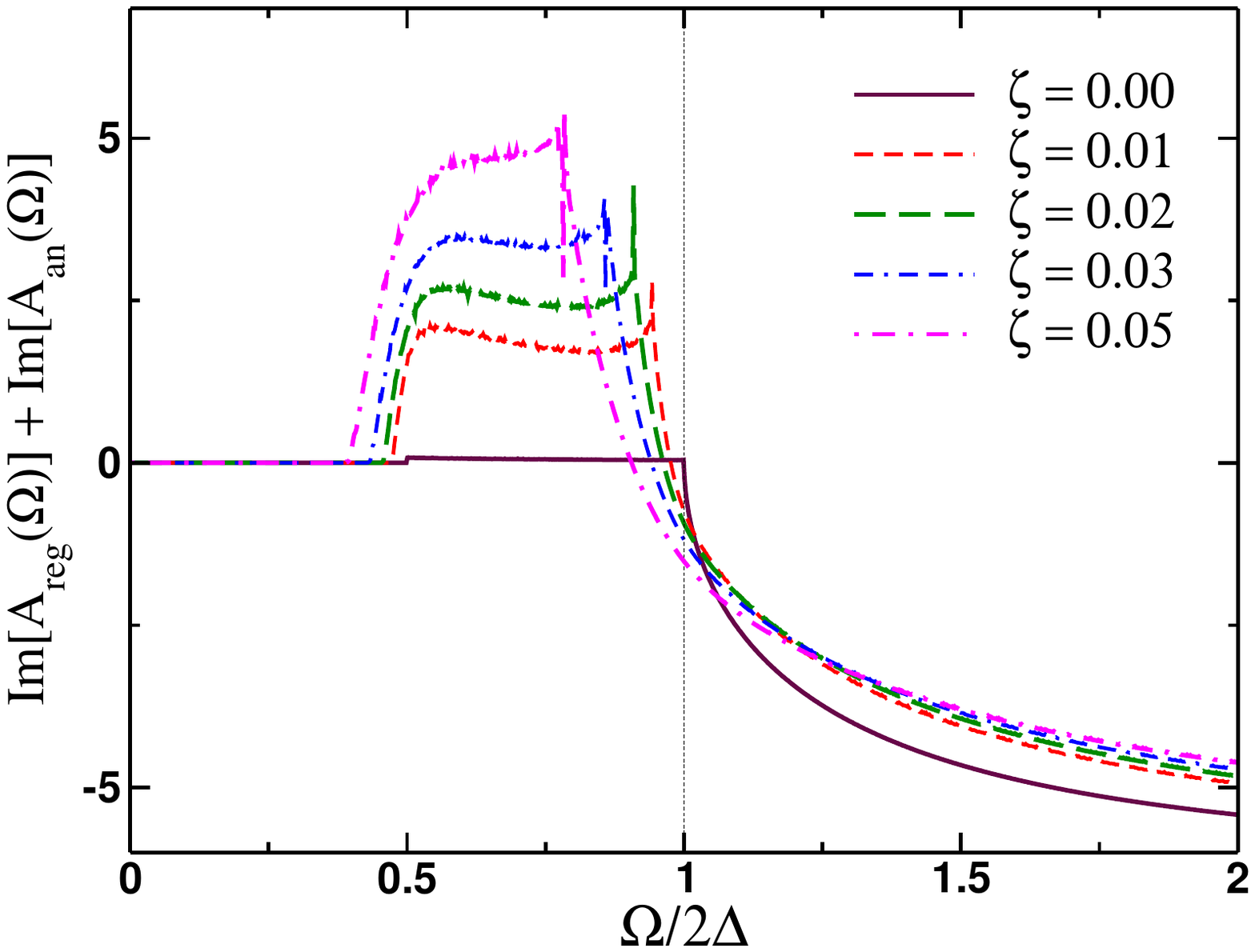}
\caption{Frequency dependence of the real and imaginary part of the function $A(\Omega)=A_{\textrm{reg}}(\Omega)+A_{\textrm{an}}(\Omega)$. The frequency is shown in the units of the pairing amplitude $\Delta$ for various values of the dimensionless parameter $\zeta=1/\tau_s\Delta$.}
\label{FigADenom}
\end{figure*}
%%%%%%%%%%%%%%%%%%%%%%%%%%%%%%

%%%%%%%%%%%%% Fig: Breg+Ban %%%%%%%%%%%%%%%%%
\begin{figure*}[t]
\includegraphics[width=0.475\linewidth]{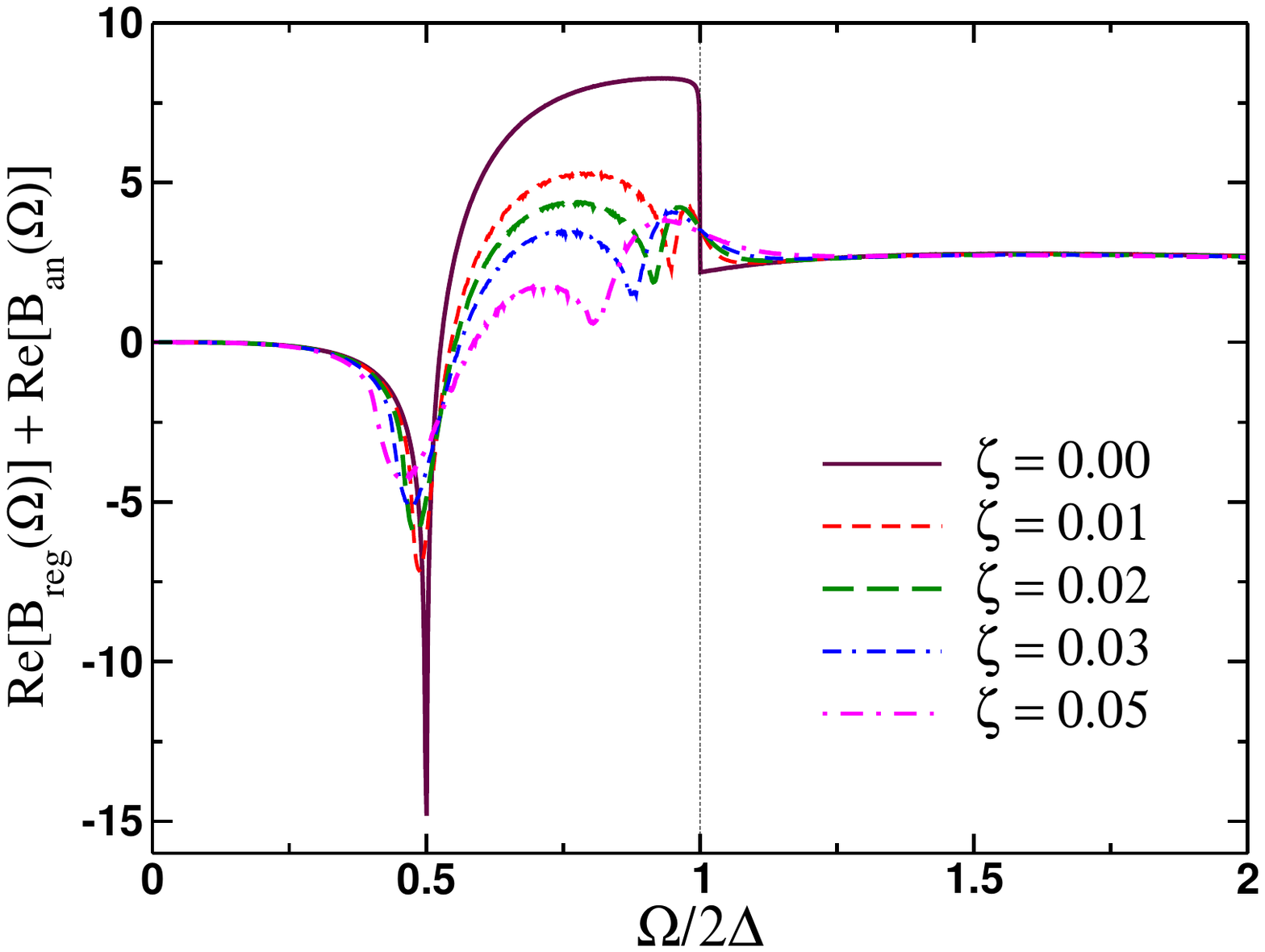}
\includegraphics[width=0.475\linewidth]{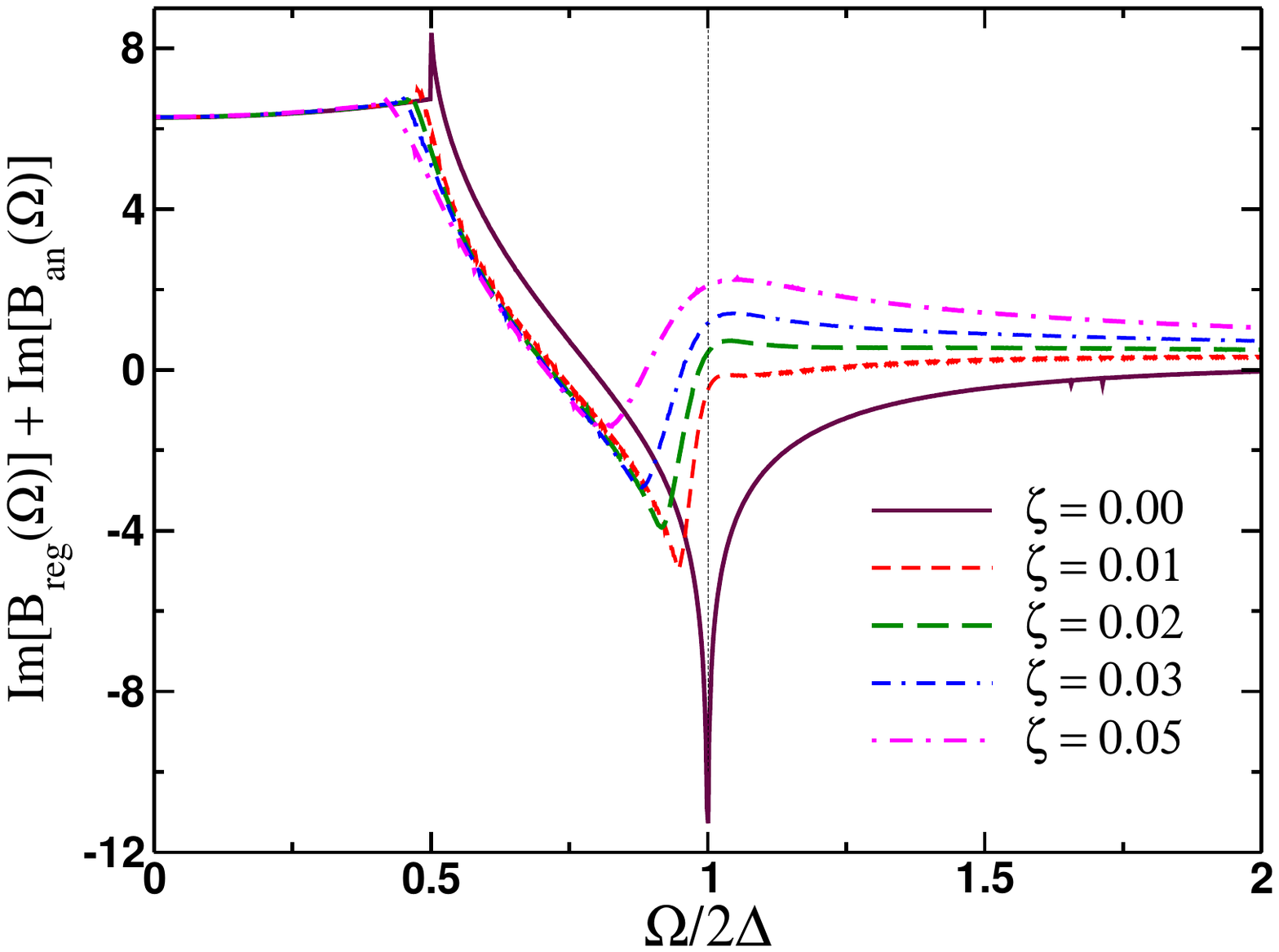}
\caption{Frequency dependence of the real and imaginary part of the function $B(\Omega)=B_{\textrm{reg}}(\Omega)+B_{\textrm{an}}(\Omega)$. The frequency is shown in the units of the pairing amplitude $\Delta$ for various values of the dimensionless parameter $\zeta=1/\tau_s\Delta$.}
\label{FigBNumer}
\end{figure*}
%%%%%%%%%%%%%%%%%%%%%%%%%%%%%%

\section{Green's functions in the Matsubara representation}
In this Section we provide the expressions for the Green's functions in the Matsubara representation. After making the substitution $\eps\to i\omega_l$, it follows that $g_\eps^{R(A)}\to g_{\omega_l}$, $g_{\eps+\Omega}^{R(A)}\to g_{\omega_l-i\Omega}$, $f_\eps^{R(A)}\to if_{i\omega_l}$ and $f_{\eps+\Omega}^{R(A)}\to if_{\omega_l-i\Omega}$. Introducing 
\beg\label{tildewldl}
\tilde{\omega}_l=\omega_l+\frac{g_{\omega_l}}{2\tau_s}, \quad \tilde{\Delta}_l=\Delta-\frac{f_{\omega_l}}{2\tau_s}
\en
and $u_{\omega_l}=\tilde{\omega}_l/\tilde{\Delta}_l$, from the normalization condition $g_{\omega_l}^2+f_{\omega_l}^2=1$ we find $g_{\omega_l}=u_{\omega_l}/\sqrt{u_{\omega_l}^2+1}$, $f_{\omega_l}=1/\sqrt{u_{\omega_l}^2+1}$. 
Function $u_{\omega_l}$ is found by solving the nonlinear equation \cite{AG1961}:
\beg\label{Eq4ul}
\left(1-\frac{1}{\tau_s\Delta}\frac{1}{\sqrt{u_{\omega_l}^2+1}}\right)u_{\omega_l}=\frac{\omega_l}{\Delta}.
\en
We note that $u_{\omega_l}=-u_{-\omega_l}$. Using this property we can now cast the expressions for the functions $A_{\textrm{reg}}(\Omega)$ and $B_{\textrm{reg}}(\Omega)$ into the following form:
\beg\label{FinAregBreg}
\begin{split}
A_{\textrm{reg}}(\Omega)&=8\pi T\sum\limits_{l=0}^{\infty}\left(\frac{1+g_{\omega_l}g_{\omega_l+i\Omega}-f_{\omega_l}f_{\omega_l+i\Omega}}{\zeta_{\omega_l}+\zeta_{\omega_l+i\Omega}}-\frac{f_{\omega_l}}{\Delta}\right), \\
B_{\textrm{reg}}(\Omega)&=8\pi iT\sum\limits_{l=0}^{\infty}\left(\frac{g_{\omega_l}+g_{\omega_l+i\Omega}}{\zeta_{\omega_l}+\zeta_{\omega_l+i\Omega}}\right)(g_{\omega_l}f_{\omega_l+i\Omega}+f_{\omega_l}g_{\omega_l+i\Omega}).
\end{split}
\en
When temperatures are close to absolute zero, we can convert the summation over $l$ to integration 
\beg\label{SumToInt}
2\pi T\sum\limits_{l=0}^\infty\to\int\limits_{0}^\infty d\omega_l.
\en
The results of the numerical calculations of the functions $A_{\textrm{reg}}(\Omega)$ and $B_{\textrm{reg}}(\Omega)$ are shown in Figs. \ref{FigADenom} and \ref{FigBNumer}. Note that the cusp like feature when $\zeta=0$ in both real and imaginary parts at $\Omega\approx2\Delta$ moves to the smaller values 
of $\Omega/2\Delta$ with an increase in the values of $\zeta$. 

\section{Electromagnetic field response kernel}
The expression for the current can be derived from the same effective action of the nonlinear $\sigma$-model, which was used to derive the Usadel equation (\ref{Eq1Usadel}). Following the avenue of Refs. \cite{Kamenev2009,Kamenev2011} by varying the corresponding part of the effective action with respect to the quantum component of the gradient vector potential ${\mathbf A}^{(q)}(t)$ for the electric current we find
\beg\label{EqCurrent}
{\mathbf j}(t)=\frac{i\pi\sigma_D}{4}\int\limits_{-\infty}^\infty dt_1\textrm{Tr}\left\{\hat{G}^R(t,t_1)\hat{\tilde{G}}^K(t_1,t)+\hat{G}^K(t,t_1)\hat{\tilde{G}}^A(t_1,t)\right\}{\mathbf A}(t_1).
\en
Here we introduced the compact notation $\hat{\tilde{G}}(t,t')=\hat{\Xi}_3\hat{G}(t,t')\hat{\Xi}_3$, $\sigma_D=2e^2\nu_0D$ is the Drude conductivity and at the intermediate stages of the calculation we have used the normalization condition (\ref{Norm}). Performing the Fourier transformation in (\ref{EqCurrent}) yields:
\beg\label{jnu}
{\mathbf j}(\omega)=-{\cal Q}(\omega,\omega'){\mathbf A}_{\omega'}, 
\en
where the kernel ${\cal Q}(\omega,\omega')$ is determined from
\beg\label{Qnnp}
\begin{split}
{\cal Q}(\omega,\omega')=\frac{\pi\sigma_D}{4i}\int\limits_{-\infty}^\infty \frac{d\eps}{2\pi}\int\limits_{-\infty}^\infty\frac{d\eps'}{2\pi}
&\textrm{Tr}\left\{\hat{\tilde{G}}^R(\eps,\eps'-\omega')\hat{{G}}^K(\eps',\eps+\omega)\right.\\&\left.+\hat{G}^K(\eps,\eps'-\omega')\hat{\tilde{G}}^A(\eps',\eps+\omega)\right\},
\end{split}
\en
which coincides with Eq. (\ref{QnnpMain}) in the main text. 
Taking into account (\ref{Gcorr}) and (\ref{Efield}), it is clear that there are many contributions to the kernel. For example, if we were to limit ourselves to the linear approximation, than using the first term in (\ref{Gcorr}) we readily obtain
\beg\label{Fominov1}
{\cal Q}(\omega,\omega')=-2\pi\delta(\omega-\omega')\left[\frac{i\sigma_D}{8}\int\limits_{-\infty}^\infty dE\textrm{Tr}\left\{
\hat{\tilde{{\cal G}}}_E^R\hat{{\cal G}}_{E+\omega}^K+\hat{{\cal G}}_E^K\hat{\tilde{{\cal G}}}_{E+\omega}^A\right\}\right],
\en
which matches the corresponding expression in \cite{Fominov2011}.

As it is stated in the main text, we will focus on computing the third harmonic contribution to the current. This means that in expression (\ref{Qnnp}) we need to single out the contributions which contain terms linear in $\check{g}_1(\eps,\eps')$:
\beg\label{QTH}
\begin{split}
{\cal Q}_3(\omega,\omega')=\frac{\sigma_D}{8i}\int\limits_{-\infty}^\infty{dE}
&\textrm{Tr}\left\{
\hat{\tilde{{\cal G}}}_{E}^R \hat{g}_1^K(E+\omega',E+\omega)+\hat{g}_1^R(E,E+\omega-\omega')\hat{\tilde{{\cal G}}}_{E+\omega}^K
\right.\\&\left.+\hat{g}_1^K(E,E+\omega-\omega')\hat{\tilde{{\cal G}}}_{E+\omega}^A+\hat{\tilde{{\cal G}}}_E^K\hat{g}_1^A(E+\omega',E+\omega)\right\}.
\end{split}
\en
At this point it proves convenient to change the integration variable from $E$ to $\eps=E+\omega'$ in the first and the fourth terms under the integral, 
so that all functions $\hat{g}_1$ have the same arguments:
\beg\label{QTH1}
\begin{split}
{\cal Q}_3(\omega,\omega')=\frac{\sigma_D}{8i}\int\limits_{-\infty}^\infty{dE}
&\textrm{Tr}\left\{
\hat{\tilde{{\cal G}}}_{E-\omega'}^R \hat{g}_1^K(E,E+\omega-\omega')+\hat{g}_1^R(E,E+\omega-\omega')\hat{\tilde{{\cal G}}}_{E+\omega}^K
\right.\\&\left.+\hat{g}_1^K(E,E+\omega-\omega')\hat{\tilde{{\cal G}}}_{E+\omega}^A+\hat{\tilde{{\cal G}}}_{E-\omega'}^K\hat{g}_1^A(E,E+\omega)\right\}.
\end{split}
\en
Using the ansatz (\ref{g1KAnsatz}) we can immediately separate the contribution which contains function $\hat{g}_{1,\textrm{an}}^K$ and then that term defines ${\cal Q}_3^{\textrm{(an)}}(\omega,\omega')$, Eq. (\ref{ThreeQsAn}), in the main text. On the second step we also separate the terms which contain the products of retarded (advanced) Green's functions, which defines ${\cal Q}_3^{\textrm{(reg1)}}(\omega,\omega')$, Eq. (\ref{ThreeQsReg}). Finally, the remaining term contains the products of advanced and retarded Green's functions, ${\cal Q}_3^{\textrm{(reg2)}}(\omega,\omega')$.

\section{Third harmonic contribution to the current}\label{Q3rd}
In this Section we will provide the expressions for the third harmonic contribution of the response kernel ${\cal Q}_3(2\Omega,\omega_{\textrm{p}})$.
\subsection{Anomalous contribution}
We start with the anomalous contribution ${\cal Q}_{3}^{(\textrm{an})}$, which is given by the sum of the following two functions:
\beg\label{Q3DLTQ}
\begin{split}
{\cal Q}_{3,Q}^{(\textrm{an})}(\Omega,\omega_{\textrm{p}})&=\frac{\pi\sigma_D}{4i}\int\limits_{-\infty}^\infty\frac{i\delta W_QdE}{\zeta_{E}^R+\zeta_{E+2\Omega}^A}\textrm{Tr}\left\{\left[\left(\hat{\tilde{{\cal G}}}_{E+\Omega}^R-\hat{\cal G}_E^R\hat{\tilde{{\cal G}}}_{E+\Omega}^R\hat{{{\cal G}}}_{E+2\Omega}^A\right)(n_{E+\Omega}-n_{E+2\Omega})\right.\right.\\&\left.\left.+\left(\hat{\tilde{{\cal G}}}_{E+\Omega}^A-\hat{\cal G}_E^R\hat{\tilde{{\cal G}}}_{E+\Omega}^A\hat{{{\cal G}}}_{E+2\Omega}^A\right)(n_{E}-n_{E+\Omega})\right]
\left(\hat{\tilde{{\cal G}}}_{E-\omega_{\textrm{p}}}^R+\hat{\tilde{{\cal G}}}_{E+2\Omega+\omega_{\textrm{p}}}^A\right)\right\}, \\
{\cal Q}_{3,\Delta}^{(\textrm{an})}(\Omega,\omega_{\textrm{p}})&=\frac{\pi\sigma_D}{4i}\Delta_1(2\Omega)\int\limits_{-\infty}^\infty\frac{(n_{E}-n_{E+2\Omega})}{\zeta_{E}^R+\zeta_{E+2\Omega}^A}\\&\times\textrm{Tr}\left\{\left(\hat{\Xi}_2-\hat{\cal G}_E^R\hat{\Xi}_2\hat{\cal G}_{E+2\Omega}^A\right)
\left(\hat{\tilde{{\cal G}}}_{E-\omega_{\textrm{p}}}^R+\hat{\tilde{{\cal G}}}_{E+2\Omega+\omega_{\textrm{p}}}^A\right)\right\}dE.
\end{split}
\en
Note that the second contribution has a pre-factor $\Delta_1(2\Omega)/\delta W_Q\gg 1$, Fig. \ref{FigABSDLT1}, and therefore we may expect that 
${\cal Q}_{3,\Delta}^{(\textrm{an})}$ will significantly exceed ${\cal Q}_{3,Q}^{(\textrm{an})}$ in a range of frequencies when $\Omega\sim\Delta$. 
The traces of the matrices entering into these expressions can be computed in a straightforward manner.
To make the expressions which appear below as compact as possible, we will use the notations $E'=E-\omega_{\textrm{p}}$ and $E''=E+2\Omega+\omega_{\textrm{p}}$.
Recall that $\hat{\Xi}_2\hat{\Xi}_2=-\hat{{\mathbbm{1}}}$ and $\hat{\Xi}_3\hat{\Xi}_2=\hat{\Xi}_1$. It then follows
\beg\label{Q3anD}
\begin{split}
&{\cal Q}_{3,\Delta}^{(\textrm{an})}(\Omega,\omega_{\textrm{p}})={i\pi\sigma_D}\Delta_1(2\Omega)\int\limits_{-\infty}^\infty\frac{(n_{E+2\Omega}-n_E)}{\zeta_{E}^R+\zeta_{E+2\Omega}^A}\left\{\left(g_E^Rf_{E+2\Omega}^A+f_E^Rg_{E+2\Omega}^A\right)\left(g_{E'}^R+g_{E''}^A\right)
\right.\\&\left.+\left(g_E^Rg_{E+2\Omega}^A+f_E^Rf_{E+2\Omega}^A+1\right)\left(f_{E'}^R+f_{E''}^A\right)\right\}dE.
\end{split}
\en
Similar although much lengthier expression is found for ${\cal Q}_{3,Q}^{(\textrm{an})}$:
\beg\label{Q3anQ}
\begin{split}
&{\cal Q}_{3,Q}^{(\textrm{an})}(\Omega,\omega_{\textrm{p}})=-{\pi\sigma_D}\delta W_Q\int\limits_{-\infty}^\infty\frac{(n_{E+\Omega}-n_{E+2\Omega})}{\zeta_{E}^R+\zeta_{E+2\Omega}^A}\left\{G_{E'E''}^{RA}\left[f_{E+\Omega}^R\left(g_{E}^Rf_{E+2\Omega}^A+f_{E}^Rg_{E+2\Omega}^A\right)
\right.\right.\\&\left.\left.+g_{E+\Omega}^R\left(g_{E}^Rg_{E+2\Omega}^A+f_{E}^Rf_{E+2\Omega}^A-1\right)\right]+F_{E'E''}^{RA}\left[
f_{E+\Omega}^R\left(g_{E}^Rg_{E+2\Omega}^A+f_{E}^Rf_{E+2\Omega}^A+1\right)\right.\right.\\&\left.\left.+g_{E+\Omega}^R
\left(g_{E}^Rf_{E+2\Omega}^A+f_{E}^Rg_{E+2\Omega}^A\right)
\right]\right\}dE-{\pi\sigma_D}\delta W_Q\int\limits_{-\infty}^\infty\frac{(n_{E}-n_{E+\Omega})}{\zeta_{E}^R+\zeta_{E+2\Omega}^A}
\left\{G_{E'E''}^{RA}\left[f_{E+\Omega}^A\right.\right.\\&\left.\left.\times\left(g_{E}^Rf_{E+2\Omega}^A+f_{E}^Rg_{E+2\Omega}^A\right)
+g_{E+\Omega}^A\left(g_{E}^Rg_{E+2\Omega}^A+f_{E}^Rf_{E+2\Omega}^A-1\right)\right]+F_{E'E''}^{RA}\left[
f_{E+\Omega}^A\right.\right.\\&\left.\left.\times\left(g_{E}^Rg_{E+2\Omega}^A+f_{E}^Rf_{E+2\Omega}^A+1\right)+g_{E+\Omega}^A
\left(g_{E}^Rf_{E+2\Omega}^A+f_{E}^Rg_{E+2\Omega}^A\right)
\right]\right\}dE
\end{split}
\en
Here we are using the shorthand notations $G_{E'E''}^{RA}=g_{E'}^R+g_{E''}^{A}$ and $F_{E'E''}^{RA}=f_{E'}^R+f_{E''}^{A}$. At temperatures close to absolute zero in order to simplify the numerical calculations we can replace 
$n_{E+\Omega}-n_{E+2\Omega}\approx2[\vartheta(-E-2\Omega)-\vartheta(-E-\Omega)]$ (here $\vartheta(x)$ is the step function), which for positive values of $\Omega$ is nonzero only when $E\in[-2\Omega,-\Omega]$. Analogously, second integral in (\ref{Q3anQ}) is nonzero for $E\in[-\Omega,0]$. We use these expressions to compute  the dependence of  ${\cal Q}_{3}^{(\textrm{an})}(\Omega,\omega_{\textrm{p}})$ on $\omega_{\textrm{p}}$ for fixed $\Omega$.

%%%%%%%%%%%%% Fig: Q3-an-DQ %%%%%%%%%%%%%%%%%
\begin{figure*}[t]
\includegraphics[width=0.475\linewidth]{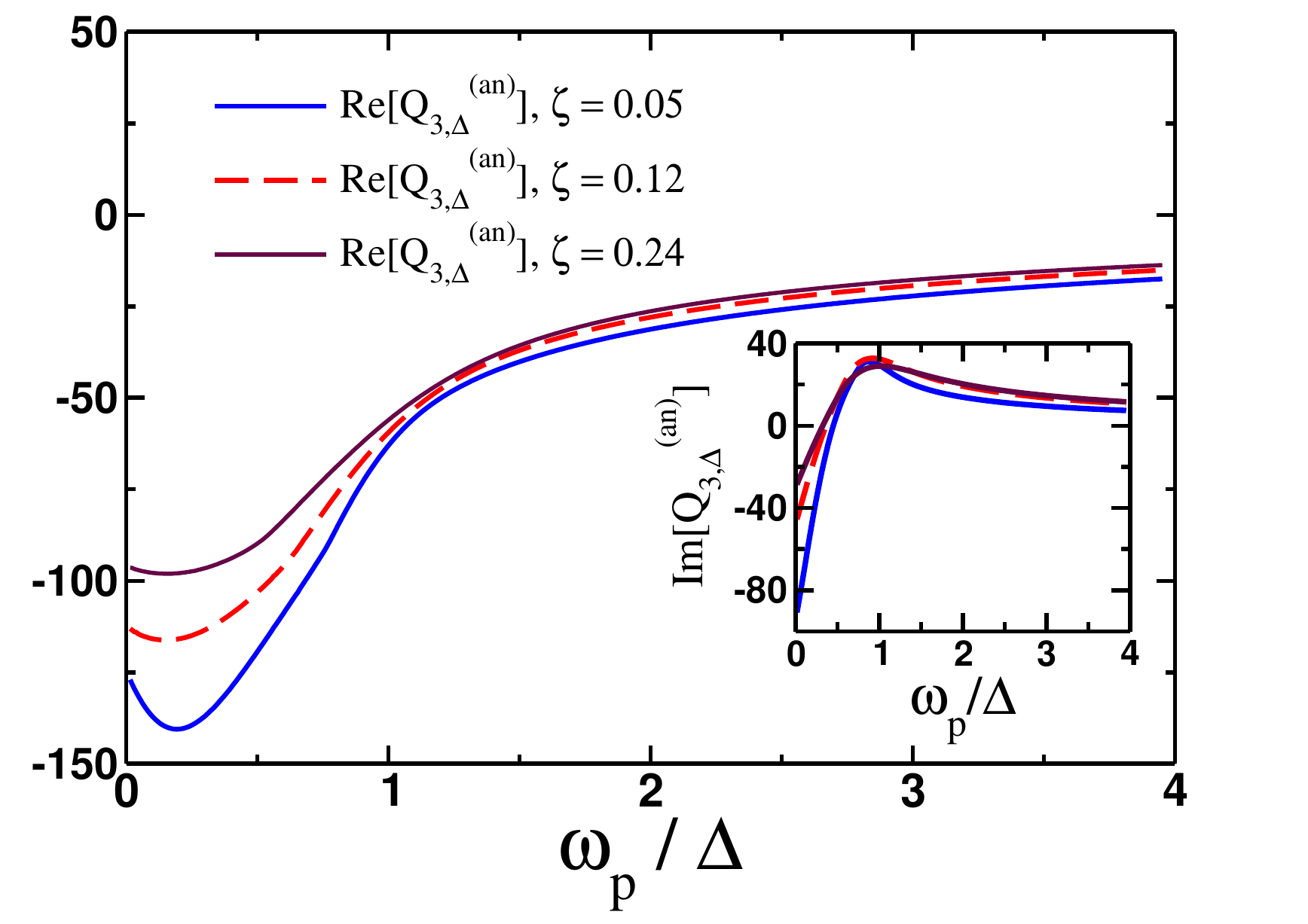}
\includegraphics[width=0.475\linewidth]{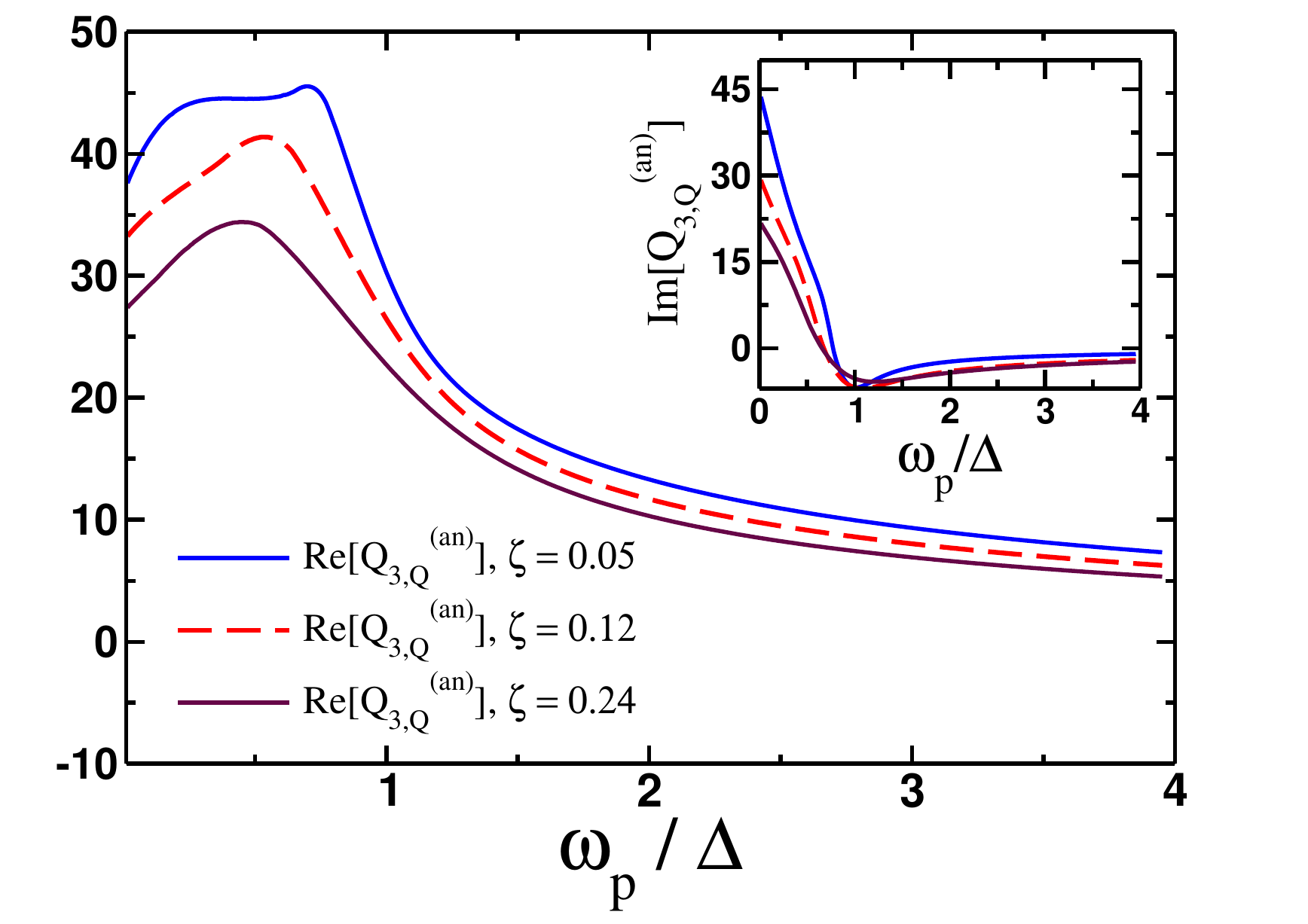}
\caption{Plots of the real (main) and imaginary (inset) parts of the functions ${\cal Q}_{3,\Delta}^{(\textrm{an})}(\Omega,\omega_{\textrm{p}})$ (left panel) and ${\cal Q}_{3,Q}^{(\textrm{an})}(\Omega,\omega_{\textrm{p}})$ (right panel) as a function of the probe frequency with the value of the pump frequency fixed to $\Omega=0.895\Delta$ for various values of the dimensionless parameter $\zeta$. Both functions are given in the units of $\delta W_Q\sigma_D$. Note that function ${\cal Q}_{3,Q}^{(\textrm{an})}$ reaches its maximum values at $\omega_{\textrm{p}}\approx\Delta_{\textrm{th}}$.}
\label{Fig-Q3an}
\end{figure*}
%%%%%%%%%%%%%%%%%%%%%%%%%%%%%%
In Fig. \ref{Fig-Q3an} we show the dependence on the probe frequency of the two terms which contribute to ${\cal Q}_3^{(\textrm{an})}={\cal Q}_{3,\Delta}^{(\textrm{an})}+{\cal Q}_{3,Q}^{(\textrm{an})}$. Given the definition (\ref{g1Kanom}) the first term ${\cal Q}_{3,\Delta}^{(\textrm{an})}$ is determined by $\hat{\rho}_\Delta$ and therefore is proportional to $\Delta_1(2\Omega)$, while the remaining term ${\cal Q}_{3,Q}^{(\textrm{an})}$ must be then proportional to $\delta W_Q$. Note that ${\cal Q}_{3,\Delta}^{(\textrm{an})}$ and ${\cal Q}_{3,Q}^{(\textrm{an})}$ enter with the opposite signs.

\subsection{Regular contribution}
The regular contribution to the kernel is given by expressions (\ref{ThreeQsReg}). We start by considering the following expression:
\beg\label{ThreeQsReg1}
\begin{split}
{\cal Q}_3^{(\textrm{reg1},R)}(\omega,\omega')&=\frac{\sigma_D}{8i}\int\limits_{-\infty}^\infty
\textrm{Tr}\left\{\hat{g}_{1}^R(E,E+\omega-\omega')\left[\hat{\tilde{{\cal G}}}_{E-\omega'}^Rn_{E+\omega-\omega'}+\hat{\tilde{{\cal G}}}_{E+\omega}^Rn_{E+\omega}\right]\right\}{dE}
\end{split}
\en
Using the definitions (\ref{DefRs}) from the main text, after some tedious algebraic manipulations similar to the ones used in derivation of (\ref{Q3anQ}) we find
\beg\label{RQRDContr}
\begin{split}
&\textrm{Tr}\left\{\hat{g}_{1}^R(E,E+2\Omega)\hat{\tilde{{\cal G}}}_{E-\omega_{\textrm{p}}}^R\right\}=-\frac{8\pi i\delta W_Q}{\zeta_E^R+\zeta_{E+2\Omega}^R}
\left\{g_{E+\Omega}^Rg_{E-\omega_{\textrm{p}}}^R\left(g_E^Rg_{E+2\Omega}^R+f_E^Rf_{E+2\Omega}^R-1\right)\right.\\&+\left.
f_{E+\Omega}^Rf_{E-\omega_{\textrm{p}}}^R\left(g_E^Rg_{E+2\Omega}^R+f_E^Rf_{E+2\Omega}^R+1\right)\right.\\&\left.+
\left(g_E^Rf_{E+2\Omega}^R+f_E^Rg_{E+2\Omega}^R\right)\left(g_{E+\Omega}^Rf_{E-\omega_{\textrm{p}}}^R+f_{E+\Omega}^Rg_{E-\omega_{\textrm{p}}}^R\right)\right\}\\&+\frac{8\pi i \Delta_1(2\Omega)}{\zeta_E^R+\zeta_{E+2\Omega}^R}\left\{f_{E-\omega_{\textrm{p}}}^R\left(g_E^Rg_{E+2\Omega}^R+f_E^Rf_{E+2\Omega}^R+1\right)+g_{E-\omega_{\textrm{p}}}^R\left(g_E^Rf_{E+2\Omega}^R+f_E^Rg_{E+2\Omega}^R\right)\right\}.
\end{split}
\en
Similar expression can be easily obtained for the second term in (\ref{ThreeQsReg1}) by replacing $g_{E-\omega_{\textrm{p}}}^R$ with 
$g_{E+2\Omega+\omega_{\textrm{p}}}^R$ and $f_{E-\omega_{\textrm{p}}}^R$ with $f_{E+2\Omega+\omega_{\textrm{p}}}^R$. Since the expression (\ref{ThreeQsReg1}) contains the functions which are analytic in the upper half-plane of the complex variable $E$, we replace the integral over $E$ with the summation over the fermionic Matsubara frequencies just like it has been done above. We repeated the same procedure for the term, which contains the advanced functions only. Finally, essentially the same type of trace as in (\ref{RQRDContr}) need to be computed in order to evaluate ${\cal Q}_3^{(\textrm{reg2},R)}(\omega,\omega')$. We will not list the resulting expression here. The dependence of these functions on $\omega_{\textrm{p}}$ is presented in Fig. \ref{Fig-Q3reg12} in the main text.

\end{appendix}

%\bibliography{higgsbiblio}

\begin{thebibliography}{10}

\bibitem{THz1}
J.~A. F\"{u}l\"{o}p, L.~P\'{a}lfalvi, G.~Alm\'{a}si, and J.~Hebling, ``Design
  of high-energy terahertz sources based on optical rectification,'' {\em Opt.
  Express}, vol.~18, pp.~12311--12327, Jun 2010.

\bibitem{THz2}
D.~N. Basov, R.~D. Averitt, D.~van~der Marel, M.~Dressel, and K.~Haule,
  ``Electrodynamics of correlated electron materials,'' {\em Rev. Mod. Phys.},
  vol.~83, pp.~471--541, Jun 2011.

\bibitem{Shimano2012}
R.~Matsunaga and R.~Shimano, ``Nonequilibrium BCS state dynamics induced by
  intense terahertz pulses in a superconducting NbN film,'' {\em Phys. Rev.
  Lett.}, vol.~109, p.~187002, 2012.

\bibitem{Shimano2013}
R.~Matsunaga, Y.~I. Hamada, K.~Makise, Y.~Uzawa, H.~Terai, Z.~Wang, and
  R.~Shimano, ``Higgs amplitude mode in the BCS superconductors
  ${\mathrm{nb}}_{1\mathrm{\text{-}}x}{\mathrm{ti}}_{x}\mathbf{N}$ induced by
  terahertz pulse excitation,'' {\em Phys. Rev. Lett.}, vol.~111, p.~057002,
  Jul 2013.

\bibitem{Shimano2014}
R.~Matsunaga, N.~Tsuji, H.~Fujita, A.~Sugioka, K.~Makise, Y.~Uzawa, H.~Terai,
  Z.~Wang, H.~Aoki, and R.~Shimano, ``Light-induced collective pseudospin
  precession resonating with Higgs mode in a superconductor,'' {\em Science},
  vol.~345, no.~6201, pp.~1145--1149, 2014.

\bibitem{THz3}
M.~Beck, I.~Rousseau, M.~Klammer, P.~Leiderer, M.~Mittendorff, S.~Winnerl,
  M.~Helm, G.~N. Gol'tsman, and J.~Demsar, ``Transient increase of the energy
  gap of superconducting NbN thin films excited by resonant narrow-band
  terahertz pulses,'' {\em Phys. Rev. Lett.}, vol.~110, p.~267003, Jun 2013.

\bibitem{Eliashberg1970}
G.~Eliashberg, ``Film superconductivity stimulated by a high-frequency field,''
  {\em Sov. Phys. - JETP Lett.}, vol.~11, p.~114, 1970.

\bibitem{Klapwijk1977}
T.~M. Klapwijk, J.~N. van~den Bergh, and J.~E. Mooij, ``Radiation-stimulated
  superconductivity,'' {\em Journal of Low Temperature Physics}, vol.~26,
  no.~3, pp.~385--405, 1977.

\bibitem{Ivlev1973}
B.~I. Ivlev, S.~G. Lisitsyn, and G.~M. Eliashberg, ``Nonequilibrium excitations
  in superconductors in high-frequency fields,'' {\em Journal of Low
  Temperature Physics}, vol.~10, no.~3, pp.~449--468, 1973.

\bibitem{VolkovKogan1973}
A.~F. Volkov and S.~M. Kogan, ``Collisionless relaxation of the energy gap in
  superconductors,'' {\em Zh. Eksp. Teor. Fiz}, vol.~65, p.~2038, 1974.
\newblock {English} translation: Sov. Phys. JETP, {\bf 38}, 1018 (1974).

\bibitem{Galaiko1972}
V.~P. Galaiko, ``Kinetic equation for relaxation processes in
  superconductors,'' {\em Sov. Phys. JETP}, vol.~34, p.~203, 1972.

\bibitem{Galperin1981}
Y.~M. Galperin, V.~I. Kozub, and B.~Z. Spivak, ``Dissipationless BCS dynamics
  with large branch imbalance,'' {\em Sov. Phys. JETP}, vol.~54, p.~1126, 1981.

\bibitem{Demsar1999}
J.~Demsar, B.~Podobnik, V.~V. Kabanov, T.~Wolf, and D.~Mihailovic,
  ``Superconducting gap $\ensuremath{\Delta}{}_{\mathit{c}}$, the pseudogap
  $\ensuremath{\Delta}{}_{\mathit{p}}$, and pair fluctuations above ${T}_{c}$
  in overdoped
  $\mathit{Y}{}_{1\ensuremath{-}\mathit{x}}\mathit{Ca}{}_{\mathit{x}}\mathit{Ba}{}_{2}\mathit{Cu}{}_{3}\mathit{O}{}_{7\ensuremath{-}\mathit{\ensuremath{\delta}}}$
  from femtosecond time-domain spectroscopy,'' {\em Phys. Rev. Lett.}, vol.~82,
  pp.~4918--4921, Jun 1999.

\bibitem{Kaindl2000}
R.~A. Kaindl, M.~Woerner, T.~Elsaesser, D.~C. Smith, J.~F. Ryan, G.~A. Farnan,
  M.~P. McCurry, and D.~G. Walmsley, ``Ultrafast mid-infrared response of
  $\textrm{YBa}_2\textrm{Cu}_3\textrm{O}_{7-\delta}$,'' {\em Science},
  vol.~287, no.~5452, pp.~470--473, 2000.

\bibitem{Averitt2001}
R.~D. Averitt, G.~Rodriguez, A.~I. Lobad, J.~L.~W. Siders, S.~A. Trugman, and
  A.~J. Taylor, ``Nonequilibrium superconductivity and quasiparticle dynamics
  in
  ${\mathrm{YBa}}_{2}{\mathrm{Cu}}_{3}{\mathrm{O}}_{7\ensuremath{-}\ensuremath{\delta}}$,''
  {\em Phys. Rev. B}, vol.~63, p.~140502, Mar 2001.

\bibitem{Leggett2005}
G.~L. Warner and A.~J. Leggett, ``Quench dynamics of a superfluid Fermi gas,''
  {\em Phys. Rev. B}, vol.~71, p.~034514, 2005.

\bibitem{Pashkin2010}
A.~Pashkin, M.~Porer, M.~Beyer, K.~W. Kim, A.~Dubroka, C.~Bernhard, X.~Yao,
  Y.~Dagan, R.~Hackl, A.~Erb, J.~Demsar, R.~Huber, and A.~Leitenstorfer,
  ``Femtosecond response of quasiparticles and phonons in superconducting
  $\textrm{YBa}_2\textrm{Cu}_3\textrm{O}_{7-\delta}$ studied by wideband
  terahertz spectroscopy,'' {\em Phys. Rev. Lett.}, vol.~105, p.~067001, Aug
  2010.

\bibitem{Beck2011}
M.~Beck, M.~Klammer, S.~Lang, P.~Leiderer, V.~V. Kabanov, G.~N. Gol'tsman, and
  J.~Demsar, ``Energy-gap dynamics of superconducting NbN thin films studied by
  time-resolved terahertz spectroscopy,'' {\em Phys. Rev. Lett.}, vol.~107,
  p.~177007, Oct 2011.

\bibitem{Eckern1979}
U.~Eckern, A.~Schmid, M.~Schmutz, and G.~Sch{\"o}n, ``Stability of
  superconducting states out of thermal equilibrium,'' {\em Journal of Low
  Temperature Physics}, vol.~36, no.~5, pp.~643--687, 1979.

\bibitem{Varma1982}
P.~B. Littlewood and C.~M. Varma, ``Amplitude collective modes in
  superconductors and their coupling to charge-density waves,'' {\em Phys. Rev.
  B}, vol.~26, pp.~4883--4893, Nov 1982.

\bibitem{Eastham2011}
R.~T. Brierley, P.~B. Littlewood, and P.~R. Eastham, ``Amplitude-mode dynamics
  of polariton condensates,'' {\em Phys. Rev. Lett.}, vol.~107, p.~040401,
  2011.

\bibitem{Varma2014}
D.~Pekker and C.~Varma, ``Amplitude/Higgs modes in condensed matter physics,''
  {\em Annual Review of Condensed Matter Physics}, vol.~6, no.~1, pp.~269--297,
  2015.

\bibitem{Moore2017}
A.~Moor, A.~F. Volkov, and K.~B. Efetov, ``Amplitude Higgs mode and admittance
  in superconductors with a moving condensate,'' {\em Phys. Rev. Lett.},
  vol.~118, p.~047001, Jan 2017.

\bibitem{Shimano2020}
R.~Shimano and N.~Tsuji, ``Higgs mode in superconductors,'' {\em Annual Review
  of Condensed Matter Physics}, vol.~11, no.~1, pp.~103--124, 2020.

\bibitem{Spivak2004}
R.~A. Barankov, L.~S. Levitov, and B.~Z. Spivak, ``Solitons and Rabi
  oscillations in a time-dependent BCS pairing problem,'' {\em Phys. Rev.
  Lett.}, vol.~93, p.~160401, 2004.

\bibitem{Burnett2005}
M.~H. Szymanska, B.~D. Simons, and K.~Burnett, ``Dynamics of the BCS-BEC
  crossover in a degenerate Fermi gas,'' {\em Phys. Rev. Lett.}, vol.~94,
  p.~170402, 2005.

\bibitem{Enolski2005}
E.~A. Yuzbashyan, B.~L. Altshuler, V.~B. Kuznetsov, and V.~Z. Enolskii,
  ``Solution for the dynamics of the BCS and central spin problems,'' {\em J.
  Phys. A}, vol.~38, p.~7831, 2005.

\bibitem{Enolski2005a}
E.~A. Yuzbashyan, B.~L. Altshuler, V.~B. Kuznetsov, and V.~Z. Enolskii,
  ``Nonequilibrium cooper pairing in the nonadiabatic regime,'' {\em Phys. Rev.
  B}, vol.~72, p.~220503(R), 2005.

\bibitem{Altshuler2005}
E.~A. Yuzbashyan, V.~B. Kuznetsov, and B.~L. Altshuler, ``Integrable dynamics
  of coupled Fermi-Bose condensates,'' {\em Phys. Rev. B}, vol.~72, p.~144524,
  2005.

\bibitem{Levitov2006}
R.~A. Barankov and L.~S. Levitov, ``Dynamical selection in developing fermionic
  pairing,'' {\em Phys. Rev. A}, vol.~73, p.~033614, 2006.

\bibitem{Landau1946}
L.~D. Landau, ``On the vibrations in the electronic plasma,'' {\em J. Phys.
  (USSR)}, vol.~10, p.~25, 1946.

\bibitem{Kadomtsev1968}
B.~B. Kadomtsev, ``Landau damping and echo in a plasma,'' {\em Soviet Physics -
  Uspekhi}, vol.~11, p.~328, 1968.

\bibitem{Gurarie2009}
V.~Gurarie, ``Nonequilibrium dynamics of weakly and strongly paired
  superconductors,'' {\em Phys. Rev. Lett.}, vol.~103, p.~075301, Aug 2009.

\bibitem{QReview2015}
E.~A. Yuzbashyan, M.~Dzero, V.~Gurarie, and M.~S. Foster, ``Quantum quench
  phase diagrams of an $s$-wave BCS-BEC condensate,'' {\em Phys. Rev. A},
  vol.~91, p.~033628, Mar 2015.

\bibitem{Yuzbashyan2006}
E.~A. Yuzbashyan, O.~Tsyplyatyev, and B.~L. Altshuler, ``Relaxation and
  persistent oscillations of the order parameter in the non-stationary BCS
  theory,'' {\em Phys. Rev. Lett.}, vol.~96, p.~097005, 2006.
\newblock Erratum: Phys. Rev. Lett. {\bf 96}, 179905 (2006).

\bibitem{Yuzbashyan2008}
E.~A. Yuzbashyan, ``Normal and anomalous solitons in the theory of dynamical
 Cooper pairing,'' {\em Phys. Rev. B}, vol.~78, p.~184507, 2008.

\bibitem{Eremin2023}
P.~Derendorf, A.~F. Volkov, and I.~Eremin, ``Nonlinear response of diffusive
  superconductors to ac-electromagnetic fields,'' {\em arXiv:2308.00838}, 2023.

\bibitem{SectionV}
For details on this issue we would like to refer the reader to the discussion in
 Section V of Ref. \cite{QReview2015}.

\bibitem{Papenkort2007}
T.~Papenkort, V.~M. Axt, and T.~Kuhn, ``Coherent dynamics and pump-probe
  spectra of BCS superconductors,'' {\em Phys. Rev. B}, vol.~76, p.~224522, Dec
  2007.

\bibitem{Axt2009}
T.~Papenkort, T.~Kuhn, and V.~M. Axt, ``Nonequilibrium dynamics and coherent
  control of BCS superconductors driven by ultrashort THz pulses,'' {\em
  Journal of Physics}, vol.~193, p.~012050, 2009.

\bibitem{Manske2014}
H.~Krull, D.~Manske, G.~S. Uhrig, and A.~P. Schnyder, ``Signatures of
  nonadiabatic BCS state dynamics in pump-probe conductivity,'' {\em Phys. Rev.
  B}, vol.~90, p.~014515, Jul 2014.

\bibitem{Aoki2015}
N.~Tsuji and H.~Aoki, ``Theory of Anderson pseudospin resonance with Higgs mode
  in superconductors,'' {\em Phys. Rev. B}, vol.~92, p.~064508, Aug 2015.

\bibitem{Kemper2015}
A.~F. Kemper, M.~A. Sentef, B.~Moritz, J.~K. Freericks, and T.~P. Devereaux,
  ``Direct observation of Higgs mode oscillations in the pump-probe
  photoemission spectra of electron-phonon mediated superconductors,'' {\em
  Phys. Rev. B}, vol.~92, p.~224517, Dec 2015.

\bibitem{Cea2016}
T.~Cea, C.~Castellani, and L.~Benfatto, ``Nonlinear optical effects and
  third-harmonic generation in superconductors: Cooper pairs versus Higgs mode
  contribution,'' {\em Phys. Rev. B}, vol.~93, p.~180507, May 2016.

\bibitem{Foster2017}
Y.-Z. Chou, Y.~Liao, and M.~S. Foster, ``Twisting Anderson pseudospins with
  light: Quench dynamics in terahertz-pumped BCS superconductors,'' {\em Phys.
  Rev. B}, vol.~95, p.~104507, Mar 2017.

\bibitem{AndersonTheorem}
P.~W. Anderson, ``Knight shift in superconductors,'' {\em Phys. Rev. Lett.},
  vol.~3, pp.~325--326, Oct 1959.

\bibitem{Silaev2019-Disorder}
M.~Silaev, ``Nonlinear electromagnetic response and Higgs-mode excitation in
  BCS superconductors with impurities,'' {\em Phys. Rev. B}, vol.~99,
  p.~224511, Jun 2019.

\bibitem{Seibold2021-Disorder}
G.~Seibold, M.~Udina, C.~Castellani, and L.~Benfatto, ``Third harmonic
  generation from collective modes in disordered superconductors,'' {\em Phys.
  Rev. B}, vol.~103, p.~014512, Jan 2021.

\bibitem{Haenel2021-Disorder}
R.~Haenel, P.~Froese, D.~Manske, and L.~Schwarz, ``Time-resolved optical
  conductivity and Higgs oscillations in two-band dirty superconductors,'' {\em
  Phys. Rev. B}, vol.~104, p.~134504, Oct 2021.

\bibitem{Yang2022-Disorder}
F.~Yang and M.~W. Wu, ``Impurity scattering in superconductors revisited:
  Diagrammatic formulation of the supercurrent-supercurrent correlation and
  Higgs-mode damping,'' {\em Phys. Rev. B}, vol.~106, p.~144509, Oct 2022.

\bibitem{LO-model}
A.~I. Larkin and Y.~N. Ovchinnikov, ``Density of states in inhomogeneous
  superconductors,'' {\em Sov. Phys. - JETP}, vol.~34, p.~1144, 1972.

\bibitem{Dzero2021}
M.~Dzero and A.~Levchenko, ``Spatially inhomogeneous magnetic
  superconductors,'' {\em Phys. Rev. B}, vol.~104, p.~L020508, Jul 2021.

\bibitem{Sherman2015-Disorder}
D.~Sherman, U.~S. Pracht, B.~Gorshunov, S.~Poran, J.~Jesudasan, M.~Chand,
  P.~Raychaudhuri, M.~Swanson, N.~Trivedi, A.~Auerbach, M.~Scheffler,
  A.~Frydman, and M.~Dressel, ``The Higgs mode in disordered superconductors
  close to a quantum phase transition,'' {\em Nature Physics}, vol.~11, no.~2,
  pp.~188--192, 2015.

\bibitem{KapitulnikRMP19}
A.~Kapitulnik, S.~A. Kivelson, and B.~Spivak, ``Colloquium: Anomalous metals:
  Failed superconductors,'' {\em Rev. Mod. Phys.}, vol.~91, p.~011002, 2019.

\bibitem{Dzero2023-SCQCP}
M.~Dzero, M.~Khodas, and Levchenko, ``Transport anomalies in multiband
  superconductors near quantum critical point,'' {\em arXiv:2308.05791}, 2023.

\bibitem{Dzero2023-Disorder}
M.~Dzero, ``Collisionless dynamics in disordered superconductors,'' {\em
  arXiv:2303.06750}, 2023.

\bibitem{AG1961}
A.~A. Abrikosov and L.~P. Gor'kov, ``Contribution to the theory of
  superconducting alloys with paramagnetic impurities,'' {\em Sov. Phys. -
  JETP}, vol.~12, p.~1243, 1961.

\bibitem{Balatsky-RMP}
A.~V. Balatsky, I.~Vekhter, and J.-X. Zhu, ``Impurity-induced states in
  conventional and unconventional superconductors,'' {\em Rev. Mod. Phys.},
  vol.~78, pp.~373--433, May 2006.

\bibitem{Austin2000}
A.~Lamacraft and B.~D. Simons, ``Tail states in a superconductor with magnetic
  impurities,'' {\em Phys. Rev. Lett.}, vol.~85, pp.~4783--4786, Nov 2000.

\bibitem{Austin2001}
A.~Lamacraft and B.~D. Simons, ``Superconductors with magnetic impurities:
  Instantons and subgap states,'' {\em Phys. Rev. B}, vol.~64, p.~014514, Jun
  2001.

\bibitem{Fominov2011}
Y.~V. Fominov, M.~Houzet, and L.~I. Glazman, ``Surface impedance of
  superconductors with weak magnetic impurities,'' {\em Phys. Rev. B}, vol.~84,
  p.~224517, Dec 2011.

\bibitem{Feigel2013}
M.~A. Skvortsov and M.~V. Feigel’man, ``Subgap states in disordered
  superconductors,'' {\em JETP}, vol.~117, pp.~487--498, 2013.

\bibitem{Yerin-EPL2022}
Y.~Yerin, A.~A. Varlamov, and C.~Petrillo, ``Topological nature of the
  transition between the gap and the gapless superconducting states,'' {\em
  Europhysics Letters}, vol.~138, p.~16005, may 2022.

\bibitem{VityaG-2002}
V.~M. Galitski and A.~I. Larkin, ``Spin glass versus superconductivity,'' {\em
  Phys. Rev. B}, vol.~66, p.~064526, Aug 2002.

\bibitem{Levitov2007}
R.~A. Barankov and L.~S. Levitov, ``Excitation of the dissipationless Higgs
  mode in a fermionic condensate,'' {\em arXiv:0704.1292}, 2007.

\bibitem{Dzero2008}
M.~Dzero, E.~A. Yuzbashyan, and B.~L. Altshuler, ``Cooper pair turbulence in
  fermionic atom traps,'' {\em Europhys. Lett.}, vol.~85, p.~20004, 2008.

\bibitem{Cavity-Higgs}
H.~Gao, F.~Schlawin, and D.~Jaksch, ``Higgs mode stabilization by photoinduced
  long-range interactions in a superconductor,'' {\em Phys. Rev. B}, vol.~104,
  p.~L140503, Oct 2021.

\bibitem{Feigel2000}
M.~V. Feigel'man, A.~I. Larkin, and M.~A. Skvortsov, ``Keldysh action for
  disordered superconductors,'' {\em Phys. Rev. B}, vol.~61, pp.~12361--12388,
  May 2000.

\bibitem{Kamenev2009}
A.~Kamenev and A.~Levchenko, ``Keldysh technique and non-linear $\sigma$-model:
  basic principles and applications,'' {\em Advances in Physics}, vol.~58,
  no.~3, pp.~197--319, 2009.

\bibitem{Kamenev2011}
A.~Kamenev, {\em Field Theory of Non-Equilibrium Systems}.
\newblock Cambridge University Press, 2011.

\bibitem{Bergeret2022}
Y.~Lu, S.~Ilić, R.~Ojajärvi, T.~T. Heikkilä, and F.~S. Bergeret, ``Reducing
  the frequency of the Higgs mode in a helical superconductor coupled to an
  lc-circuit,'' 2022.
 
\bibitem{Marchetti2002}
F.~M. Marchetti and B.~D. Simons, ``Tail states in disordered superconductors
  with magnetic impurities: the unitarity limit,'' {\em Journal of Physics A:
  Mathematical and General}, vol.~35, p.~4201, may 2002.

\bibitem{Kharitonov2012}
M.~Kharitonov, T.~Proslier, A.~Glatz, and M.~J. Pellin, ``Surface impedance of
  superconductors with magnetic impurities,'' {\em Phys. Rev. B}, vol.~86,
  p.~024514, Jul 2012.



\end{thebibliography}

\end{document}